\documentclass[twocolumn, showpacs, jcp]{revtex4-1}% floatfix,
\usepackage{amsmath}
\usepackage{graphicx}
\usepackage{float}
\usepackage{comment}
\usepackage{color}
\usepackage{soul}

\begin{document}
\title{A scale-bridging modeling approach for anisotropic organic molecules at patterned semiconductor surfaces}
\author{Nicola Kleppmann and Sabine H. L. Klapp}
\affiliation{
Institut f\"ur Theoretische Physik,
Technische Universit\"at Berlin,
Hardenbergstr. 36,
10623 Berlin,
Germany
}
\date{\today}
\begin{abstract}
Hybrid systems consisting of organic molecules at 
inorganic semiconductor surfaces are gaining increasing importance 
as thin film devices for optoelectronics. The efficiency of such devices
strongly depends on the collective behavior of the adsorbed molecules. 
In the present paper we propose a novel, coarse-grained model addressing the
condensed phases of a representative hybrid system, that is, para-sexiphenyl 
(6P) at zinc-oxide (ZnO).  Within our model, intermolecular interactions are represented via
a Gay-Berne potential (describing steric and van-der-Waals interactions)
combined with the electrostatic potential between two linear quadrupoles.
Similarly, the molecule-substrate interactions include a coupling between a linear molecular quadrupole to the electric field 
generated by the line charges characterizing ZnO(10-10). 
To validate our approach, we perform equilibrium Monte Carlo simulations, 
where the lateral positions are fixed to a 2D lattice, while the rotational degrees of freedom are continuous. We use these simulations 
to investigate orientational ordering in the condensed state.
We reproduce various experimentally observed features such as
the alignment of individual molecules with the line charges on the surface, the formation of a
standing uniaxial phase with a herringbone structure, as well as the formation of a lying nematic
phase.
\end{abstract}

\maketitle

\section{Introduction}\label{sec:intro}
Hybrid structures made of conjugated organic molecules (COMs) adsorbed at inorganic semiconductor substrates open a novel field
of application in the field of optoelectronics. On the one hand, the highly ordered crystalline and electronic structure of semiconductor surfaces leads to 
well-defined surface patterns. On the other hand, the anisotropic character of the COMs allows them to self-assemble into a variety
of structures each characterized by different features in terms of functionality (charge transfer, optimization of band-gaps, excitation transfer).
Combining these materials thus leads to tunable hybrids whose characteristic properties, such as the work-function and charge carrier mobility, 
can be tailored through the choice molecules and substrates \cite{Koch2008, Roales2012, Li2013}, as well as through the self-assembled structures formed by the adsorbed molecules \cite{Hlawacek2013, Zojer2000, Simbrunner2013}.
%Combining these materials thus leads to tunable hybrids whose characteristic properties can be tailored through both the choice of molecules and substrates 
%involved \cite{Koch2008, Roales2012, Li2013} and the self-assembled structures formed therein 
%\cite{Hlawacek2013, Zojer2000, Simbrunner2013}. 
This wide range of possibilities could not be achieved with an organic or inorganic component alone.

However, a comprehensive understanding of the complex interplay of molecular structure, substrate characteristics, and experimental conditions, and the consequences
for self-assembly of the COMs is still missing \cite{Simbrunner2011}. 
Additionally the majority of the knowledge for inorganic semiconductors \cite{Vvedensky2009, Pohl2013} is not immediately transferable to organic 
semiconductors, because of the anisotropy of organic semiconductor molecules and the vast number of different molecules with different anisotropic physiochemical properties \cite{Clancy2011}.
%This contrasts the situation for inorganic semiconductors, where growth and self-assembly are 
%well understood on both microscopic and macroscopic level for many different material systems \cite{Vvedensky2009, Pohl2013}. 

From a theoretical point of view, a major challenge in understanding self-assembled structures of hybrid systems is that
these structures involve the {\it collective} (i.e., many-particle) behavior occuring on length scales beyond those
achievable by traditional approaches such as electronic density functional theory (DFT). This calls for more coarse-grained computational approaches where,
to some extent, microscopic features are neglected in favor of relevant ``ingredients'' for collective behavior.

In this spirit, we propose in the present study a classical, coarse-grained model for a representative hybrid system, that is,
the conjugated organic molecule para-sexiphenyl (6P) adsorbed at zinc-oxide (ZnO), particularly the substrate facette ZnO(10-10). 

 Experimentally, systems of 6P have been studied both in the three-dimensional (3D) bulk crystal phase \cite{Resel2008} and in
film-like geometries (see, e.g. Refs.~\cite{Zojer2000, Hlawacek2013}
and references therein). As to the substrate, ZnO is known to have suitable energy characteristics for many opto-electronic 
applications \cite{Blumstengel2008}, and it has been studied experimentally in combination with various COMs including 6P \cite{Komolov2005, Blumstengel2010}.
Indeed, ZnO is already used in prototype
hybrid devices \cite{Huang2014,Tong2012}. Here we are particularly interested in the ZnO(10-10) facette, since the latter is
characterized by an alternating arrangements of positively charged Zn atoms and negatively charged O atoms. This leads to a stripe-like 
electrostatic surface field, which has already been found to have a strong impact
on the shape and orientational structure of adsorbed islands of 6P \cite{Blumstengel2010}.
Nevertheless, while many experimental studies of both, 6P and ZnO, exist, little is known about
the details and, more importantly, the origin of the orientational ordering, both in equilibrium
and during growth. 

So far, most theoretical studies of 6P at surfaces are based on DFT. For example,
Berkebile {\it et al.} have studied 6P on Cu(110) \cite{Berkebile2008, Berkebile2011}. They found
that the periodicity of the 6P layer has a large influence on the energetics of the resulting hybrid system, as these depend e.g. on how the $\pi$ orbitals 
of the adsorbed 6P molecules lie in relation to the hybridization states that are available at the metal surface. This observation further supports the significance of the self-assembled structures for device functionality.
Other DFT studies emphasize the influence of the molecule-substrate interactions, e.g., Braun and
Hla found that 6P molecules adsorb with alternating twist of the $\pi$ rings if physisorbed on Ag(111) \cite{Braun2005}.
Quite recently, Della Sala {\it et al.} \cite{Sala2011} studied 
the behavior of an {\it individual} 6P molecule
on a ZnO(10-10) surface. This study demonstrates that 6P adsorbs in a planar
configuration and aligns with the positively charged rows of Zn atoms on the surface. Moreover, Goose \textit{et al.} 
\cite{Goose2010} and Hlawacek \textit{et al.} \cite{Hlawacek2008} have considered the diffusion behavior of 6P at step edges.

Due to computational reasons DFT is still restricted to single molecules or very small clusters of molecules.
Therefore, various recent studies have attempted 
to investigate 6P-substrate systems by atomistic molecular dynamics (MD), where the molecules and substrates are
modelled with atomic resolution, but the dynamics is determined by 
classical force fields. An example is the study by Potocar {\it et al.} who consider the
transition from lying small clusters to the upright larger clusters of 6P molecules \cite{Potocar2011} at 6P crystalline surfaces.
We also mention recent MD simulations of 6P bulk crystals \cite{Socci1993} and organic-organic heterosurfaces
involving 6P \cite{Koller2006}.

To access even larger length and time scales than those achievable by all-atom MD, one typically employs 
Monte-Carlo (MC) simulations targeting either the equilibrium or growth behavior (kinetic MC).
However, MC approaches towards complex hybrids such as 6P/ZnO are in their infancy.
While for some problems it is sufficient to treat 6P as a point-like particle \cite{Tumbek2012}, it is well established that the mobility of
molecules is influenced by the orientational degrees of freedom  \cite{Raut1998, Raut1999}.

Within the few existing MC studies of anisotropic molecules at interfaces, the molecules' orientations
are typically strongly restricted; e.g., molecules are modelled by strings
occuping multiple lattice sites \cite{Ruiz2004, Choudhary2006, Jana2013}, 
or the orientations are restricted to a 2D plane \cite{Hopp2012}
(note that the latter study also considers the influence of a striped substrate). 
Such restrictions are inappropriate for 6P at ZnO, where it is known from experiments that the molecules
can explore the full, 3D space of orientations \cite{Blumstengel2010}.

In the present paper we thus propose a coarse-grained model of 6P at ZnO(10-10) suitable for MC simulations
with continuous, 3D orientations. Within our model, intermolecular interactions are modelled via
a Gay-Berne (GB) potential (describing steric and van-der-Waals interactions)
combined with the electrostatic potential between two linear quadrupoles.
Similarly, the molecule-substrate interactions are described via a combination of an integrated Lennard-Jones (LJ) potential
and the coupling between a linear quadrupole
and the electric field stemming from line charges characterizing ZnO(10-10) \cite{Sala2011}. To understand the interplay of the various 
contributions to the interaction Hamiltonian, we perform
 MC simulations 
for molecules that have their x- and y-coordinates 
fixed to lattice sites. The z-coordinate is determined as a function of the tilt angle between the molecule's 
long axis and the surface normal. We motivate these lattice-like simulations as follows: First, real ZnO(10-10) surfaces 
have a well-defined unit cell
characterised by a charge pattern. From DFT calculations \cite{Sala2011} and experiments \cite{Sala2011,Blumstengel2010} it is known that this electrostatic pattern 
is so strong that, at least,  the molecule's y-coordinates are {\it essentially} fixed through the stripes. Second, restricting the 
lateral positions reduces computational effort. Third, lattice-models to investigate liquid-crystalline phase behaviour are well 
established in the literature, prominent examples being the Zwanzig model (involving discretised translational motion and 
discretised rotations) \cite{Kundu2014,Fischer2009,Dickman2009,Feng2011} and the Lebwohl-Lasher model (particles fixed to lattice sites, continuous 
rotational motion) \cite{Shekhar2012,Luo2014,Ghoshal2014}. These models have been successfully used to study orientational 
ordering both in bulk \cite{Shekhar2012} and in spatially confined systems \cite{Fischer2009,Dickman2009,Feng2011,Luo2014}.

%To understand the interplay of 
%the various contributions to the interaction Hamiltonian, we perform MC simulations for molecules positionally fixed to
%lattice sites. 
Our numerical results indicate that the model is capable of displaying various structures seen in experiments.

The rest of this paper is organized as follows. In Sec.~\ref{Sec:2Model} we develop the interaction Hamiltonian defining 
our model which involves both, non-electrostatic terms (Sec.~\ref{Sec:2A_non_electrostat}) and electrostatic contributions (Sec.~\ref{Sec:2B_electrostat}).
To this end, we analyze the charge distribution of 6P in terms of multipoles. Sections~\ref{Sec:3A_simulation} and \ref{sec:3B_evaluation} describe technical details
and target quantities, respectively, of our MC simulations. Numerical results are presented in Section~\ref{Sec:3C_results}. 
We close with a brief summary and some conclusions in Sec.~\ref{Sec:4Conclusions}.

\section{Modeling 6P molecules on a patterned substrate}\label{Sec:2Model}
\begin{figure}
\includegraphics[bb=0 0 300 150]{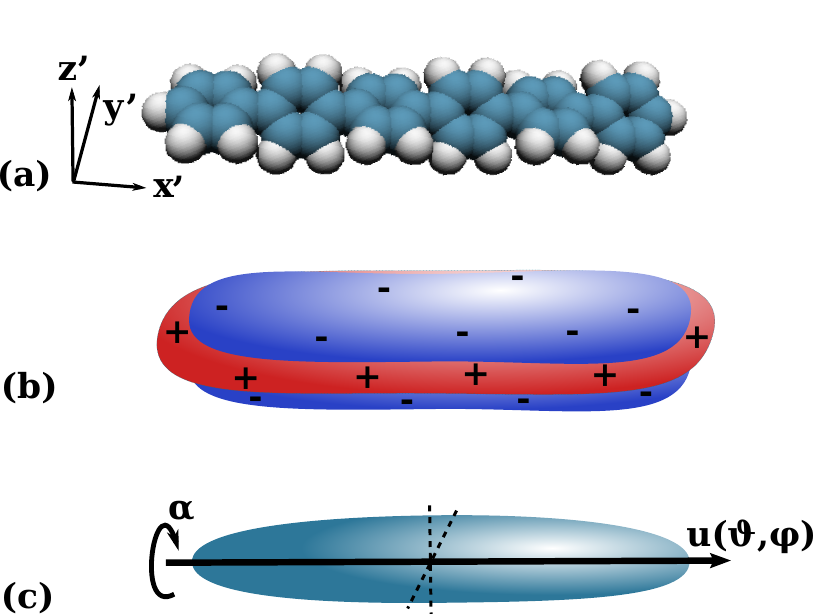}
\caption{(Color online) Three sketches of a 6P molecule with different degrees of sophistication: (a) atomistic model (twisted configuration) with the blue (white) parts representing 
C-(H-)atoms, using atomic coordinates obtained by Palczynski \textit{et al.} \cite{priv_Palczynski}, (b) roughly sketched charge
distribution with negative $\pi$-orbitals above and below the molecule and positively charged H atoms around the edge, and (c) representation as an uniaxial ellipsoid. 
In part (a) we introduce the three eigendirections of the molecules, where x$^,$ is parallel to the longest axis of the molecule and z$^,$ 
parallel to the shortest. The uniaxial ellipsoid in (c) is characterized by the vector $\textrm{\textbf{u}}(\vartheta, \varphi)$, which lies parallel to the eigendirection x$^,$
($\vartheta$, $\varphi$ are Euler angles in the space-fixed coordinate system). The ellipsoid is insensitive to rotation by the angle $\alpha$ around the axis 
$\textrm{\textbf{u}}(\vartheta, \varphi)$.}
\label{fig:1_mol_model}
\end{figure}

A system of anisotropic molecules on a patterned substrate is subject to a multitude of interactions, both between different molecules and between molecules and substrate.
In the following we aim to construct a simplified, coarse-grained Hamiltonian $H$, which still takes into account 
the most significant interactions occurring in a system of 6P molecules on a ZnO(10-10) substrate. One main simplification is that we neglect the internal atomistic structure of a 6P molecule (and likewise that of the substrate). The molecule's atomistic structure, consisting 
of C- and H-atoms, is sketched in Fig.~\ref{fig:1_mol_model}(a). Here we rather represent each molecule as a rigid body, specifically, an uniaxial ellipsoid, as sketched in 
Fig.~\ref{fig:1_mol_model}(c). However, despite this strong simplification we do take into account the fact that 6P has a complex
charge distribution, which is indicated in Fig.~\ref{fig:1_mol_model}(b). 

To proceed, we divide the (potential part of the) Hamiltonian into
\begin{equation}
H=H_{\textrm{mol-mol}}+H_{\textrm{mol-subs}}\textrm{,}\label{eqn:hamiltonian}
\end{equation}
where the subscripts ``mol'' and ``subs'' refer to the molecule and substrate, respectively.

In the following sections we will discuss these Hamiltonians in detail and analyze the implicit assumptions made. Specifically, Sec.~\ref{Sec:2A_non_electrostat} discusses the non-electrostatic 
interactions both between the molecules as well as between a molecule and the substrate, and Sec.~\ref{Sec:2B_electrostat} introduces the corresponding electrostatic 
interactions.

\subsection{Non-electrostatic interactions}\label{Sec:2A_non_electrostat}

To model the non-electrostatic part of the molecule-molecule interaction, each molecule is represented by an uniaxial ellipsoid as illustrated in Fig.~\ref{fig:1_mol_model}(c).
The role of the biaxiality [which is indeed present in true 6P molecules, see Fig.~\ref{fig:1_mol_model}(a)] will be discussed below. As molecules on the nanometer 
length scale display an attractive (van-der-Waals) interaction besides their repulsive core,
we use a generalization of the LJ potential for anisotropic molecules, that is, the GB potential \cite{Gay1981, Cleaver1996}. 
Specifically, we use the formulation suggested by Golubkov and Ren \cite{Golubkov2006}. The resulting potential of two molecules with orientations
\textbf{u}$_i$, \textbf{u}$_j$ and connection vector \textbf{r}$_{ij}=\textbf{r}_{i}-\textbf{r}_{j}$ is given by \cite{Note_GB}
\begin{align}
\notag V_{\textrm{GB}}(\textbf{u}_i, \textbf{u}_j, \textbf{r}_{ij})&=4 \epsilon(\textbf{u}_i, \textbf{u}_j, \textbf{r}_{ij})\\ 
\notag &\cdot \left(\left[\frac{\sigma_0 d_w}{r_{ij}-\sigma(\textbf{u}_i, \textbf{u}_j, \hat{\textbf{r}}_{ij})+d_w\sigma_0}\right]^{12}\right.\\ 
 &-\left.\left[\frac{\sigma_0 d_w}{r_{ij}-\sigma(\textbf{u}_i, \textbf{u}_j, \hat{\textbf{r}}_{ij})+ d_w\sigma_0}\right]^6\right) \textrm{.}\label{eq:GB}
\end{align}
The position of the potential minimum for molecules with parallel long axes is determined by $\sigma_0$, and $d_w$ 
determines the softness of the molecules. Detailed expressions for the configuration-dependent well depth $\epsilon(\textbf{u}_i, \textbf{u}_j, \textbf{r}_{ij})$ and 
contact distance $\sigma(\textbf{u}_i, \textbf{u}_j, \hat{\textbf{r}}_{ij})$ can be found in the work of Gay and Berne 
\cite{Gay1981, Cleaver1996}.
\begin{center}
\begin{figure*}
\begin{minipage}{17cm}
\includegraphics[bb=0 0 600 150]{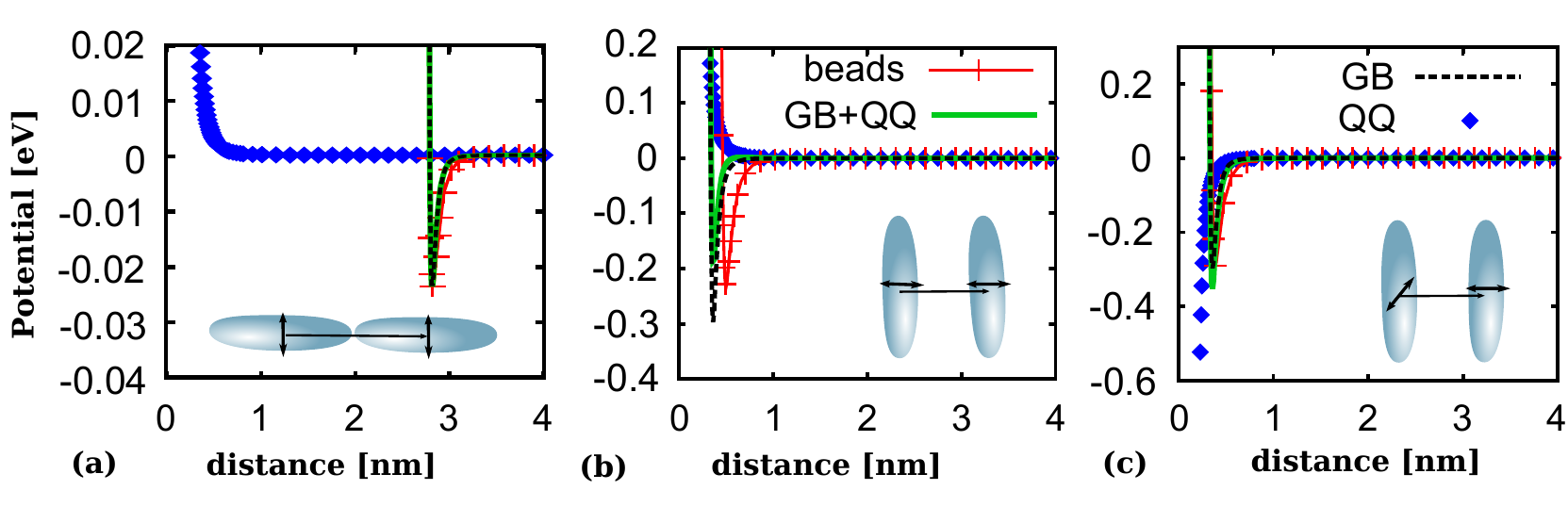}
\caption{(Color online) Parts (a)-(c) show the contributions to the molecule-molecule interaction energy for different pair configurations. Here, the legends in parts (b) and
(c) apply to the corresponding graphs all three parts. 
In the sketches in the bottom of each plot,
the molecules are shown in the representation introduced in Fig.~\ref{fig:1_mol_model}(c). The orientation of the additional 
linear quadrupole moment (see Sec.~\ref{Sec:2B_electrostat}) is indicated through the double arrow at the center of each molecules.
 Each part includes the potential between the molecules evaluated from the bead-chain parametrization suggested by Golubkov and Ren \cite{Golubkov2006} 
 as a crosshatched (red) line named `beads'. The other three traced potentials are the GB potential (GB), the quadrupole-quadrupole interaction (QQ) and the sum of the two (GB+QQ). 
 Note the differences in the energy scales.}
\label{fig:2_interactions}
\end{minipage}
\end{figure*}
\end{center}

To parametrize the GB potential in accordance to 6P we use the bead-chain  
model introduced in the work of Golubkov and Ren \cite{Golubkov2006}. These authors started from  
the GB interaction potential of two benzene 
rings, which is known. In this spirit, we can model the interaction of two planar 6P molecules through the interaction of two planar chains of benzene rings (beads). 
As a further step of simplification, we parametrize the uniaxial interaction potential 
$V_{\textrm{GB}}(\textbf{u}_i, \textbf{u}_j, \textbf{r}_{ij})$ [see Eq.~(\ref{eq:GB})] to match the potential gained from 
the interactions of the two benzene bead chains. Examples of both, the bead-chain and the GB potential are shown in Fig.~\ref{fig:2_interactions}, where we focus
on three relevant pair configurations.

The geometry of the resulting coarse-grained 6P ``molecules'' is characterized through the molecular length $l$ and diameter $d$. 
These values (as well as those of the other GB parameters) are listed in Table~\ref{tab:1GBParam}.  Moreover, the 
orientation  corresponding to the uniaxial GB ``molecule'' is defined by the vector $\textbf{u}_i=(\sin\vartheta \,\cos\varphi, \sin\vartheta \,\sin\varphi, \cos\vartheta)$ [see Fig.~\ref{fig:1_mol_model}(c)].

Next we consider the non-electrostatic part of the interaction between a molecule and the substrate. We assume that the substrate is a (smooth) plane 
located in the $x$-$y$-plane (at $z=$0) of our system. In the literature (see, e.g. \cite{Stelzer1997,Steuer2004,Micheletti2005}) several expressions for the interaction between a GB-like particle 
and a planar substrate have been proposed and used for simulations of, e.g., GB bulk fluids in contact with a wall \cite{Barmes2005} or confined GB films \cite{Gruhn1998}. 
Here we use a simpler expression which is motivated by two important results from DFT calculations \cite{Sala2011}: First, the energetically most favourable distance between the molecule's center of mass and the surface is given by $z_{\textrm{min}}\approx0.35$~nm, 
which corresponds approximately to the diameter $d$ of our GB-particles (see Table I). Second, at $z_{\textrm{min}}\approx0.35$~nm the most favourable orientation 
is planar to the surface, i.e., the (azimuthal) Euler angle $\vartheta$ of the molecule [see Fig.~\ref{fig:1_mol_model}(c)] is $\pi/2$.\\

In our simulations we realize this situation as follows. To start with, the $z$-position of each molecule is restricted to the range 
$z_{\textrm{min}}<=z_i<=z_{\textrm{min}}-d/2+l/2$. The lower limit of this interval corresponds to a planar orientation ($\vartheta=\pi/2$) , while the upper limit 
corresponds to a molecule standing upright ($\vartheta=0$). Next, we introduce a potential which, at the same time, favours the distance 
$z=z_{\textrm{min}}\approx0.35$~nm of the centre of mass and the planar orientation of the molecule's long axis. To this end we use an integrated LJ potential \cite{Hansen2006} given by
\begin{equation}
 V_{\textrm{LJ}}(\textbf{u}_i)=\epsilon_{s} \left(\frac{2}{15}\left(\frac{\sigma_{s}}{z(\textbf{u}_i)}\right)^9-\left(\frac{\sigma_{s}}{z(\textbf{u}_i)}\right)^3\right)\textrm{.}
\label{eq:V_LJ}
\end{equation}
Here, $\sigma_s$ denotes a constant length (to be defined in accordance with $z_{\textrm{min}}$, see below), and the function $z(\textbf{u}_i)$ is defined as
\begin{equation}
 z(\textbf{u}_i)=\sqrt{\left(\frac{d}{2}\right)^2(\sin\vartheta)^2+\left(\frac{l}{2}\right)^2(\cos\vartheta)^2}+\left(z_{\textrm{min}}-\frac{d}{2}\right)\textrm{.}\label{eqn:z_vartheta}
\end{equation}
By construction, $z(\textbf{u}_i)=z_{\textrm{min}}$ for $\vartheta=\pi/2$ while $z(\textbf{u}_i)= z_{\textrm{min}}+(d-l)/2 \approx 1.57$~nm for $\vartheta=0$, in accordance with the $z$-interval considered here.

To favour energetically the first case (consistent with the DFT calculations \cite{Sala2011}) we adjust the constant $\sigma_s$ 
in Eq.~(\ref{eq:V_LJ}) accordingly. The minimum of the attractive well of the potential $V(z)$ in Eq.~(\ref{eq:V_LJ}) 
is located at $z_0 = (2/5)^{1/6} \sigma_s$. Setting $z_0=z_{\textrm{min}}$ we find $\sigma_s=0.408$~nm. 
The potential depth, $\epsilon_s$, is not known and therefore an adjustable parameter. 
As an estimate we choose $\epsilon_s = 0.28$~eV. In this way, the potential plotted in Fig.~\ref{fig:6_rot_pot}(c) 
has the same order of magnitude as the molecule-substrate interaction energies found by Della Sala et al. \cite{Sala2011} for 6P on ZnO(10-10).

\begin {table}%
\begin {tabular}{c  c  c  c  c  c  c}
\hline\hline
$d_w$\,\,& \,\,$l$~[nm]\,\,&\,\, $d$~[nm]\,\, & \,\,$\epsilon_0$~[eV p.p.] \,\,& \,\,$\epsilon_e/\epsilon_s$\,\,&\,\, $\mu$ \,\,&\,\, $\nu$ \\
\hline
0.6 \,\,&\,\, 2.79\,\, & \,\,0.335\,\,& \,\,0.07\,\,& \,\, 1/12.5\,\,& \,\, 2.0 \,\, & \,\, 1.0 \\
\hline\hline
\end {tabular}
\caption {Parametrization of the inter-molecular GB potential for 6P molecules based on the bead model and parameters for individual benzene rings suggested by Golubkov 
and Ren \cite{Golubkov2006}. The abbreviation p.p. stands for `per particle'. }
\label{tab:1GBParam}
\end {table}

\subsection{Electrostatic interactions}\label{Sec:2B_electrostat}

We now turn to the electrostatic molecule-molecule and molecule-substrate interactions. These are described via appropriate multipole moments.

\subsubsection{The multipole moments of 6P molecules}\label{Sec:2B1_multipoles}

To start with, we analyze the molecular charge distribution, which is sketched in Fig.~\ref{fig:1_mol_model}(b), 
in terms of its multipole moments. 
To ensure that our model is robust we determine the moments for two different molecular configurations, a planar and a twisted molecule. 

The basic configuration used in our modeling is the planar molecule, which was also used by Della Sala \textit{et al.} \cite{Sala2011}. 
To define this configuration, we use the partial charges $q_l$ and atomic positions $\textbf{r}_{l}$ (where $l=1,\dots,M$ with $M$ being the total number
of atoms in the molecule) gained from a planar 6P model 
constructed in \mbox{\textsc{MarvinSketch}} by \mbox{ChemAxon} \cite{MarvinSketch}.
The second molecular configuration that we study is a 6P molecule with twisted benzene rings. This configuration is the relevant one for
a single 6P molecule in vacuum. The corresponding partial charges and atomistic coordinates were determined by Palczynski \textit{et al.} \cite{Palczynski2014, priv_Palczynski}. 

For reasons discussed later, the highest multipole moment considered is the
hexadecapole. The explicit expressions are given by Gray and Gubbins \cite{Gray1984}:
\begin{align}
 q&=\sum_{l=1}^{M} q_l \textrm{,}\label{Eq:Monopole}\\
 p_{\alpha}&=\sum_l q_l r_{l\alpha}\textrm{,}\label{Eq:Dipole}\\
 Q_{\alpha \beta}&=\sum_l \frac{1}{2} q_l (3r_{l\alpha} r_{l\beta} - r_l^2 \delta_{\alpha\beta})\textrm{,}\\
\notag O_{\alpha \beta \gamma} &= \sum_l \frac{1}{2} q_l(5r_{l\alpha} r_{l\beta} r_{l\gamma}\label{Eq:Quadrupole}\\\
 & \qquad-r_{l\alpha}r_l^2 \delta_{\beta \gamma}-r_{l\beta}r_l^2 \delta_{\alpha \gamma} -r_{l\gamma}r_l^2 \delta_{\alpha \beta})\textrm{,}
 \end{align}
 \begin{align}
\notag H_{\alpha \beta \gamma \eta} &= \sum_l \frac{1}{8} q_l(35r_{l\alpha} r_{l\beta} r_{l\gamma}r_{l\eta}-5r_{l\alpha}r_{l\beta}r_l^2 \delta_{\gamma \eta}\\
\notag &\qquad -5r_{l\beta}r_{l\gamma}r_l^2 \delta_{\alpha \eta}-5r_{l\gamma} r_{l\alpha}r_l^2 \delta_{\beta \eta}\\
\notag &\qquad -5r_{l\alpha}r_{l\eta}r_l^2 \delta_{\beta \gamma}-5r_{l\beta}r_{l\eta}r_l^2 \delta_{\alpha \gamma} \\
 &\qquad -5r_{l\gamma}r_{l\eta}r_l^2 \delta_{\alpha \beta}
              +\delta_{\alpha \beta}\delta_{\gamma \eta} + \delta_{\alpha \gamma}\delta_{\beta \eta}+ \delta_{\alpha \eta}\delta_{\beta \gamma})\textrm{.}\label{Eq:Hexadec}
\end{align}
In Eqs.~(\ref{Eq:Monopole})-(\ref{Eq:Hexadec}) the sums run over all partial charges $l=1,\dots,M$ in the molecule. 
The indices $\alpha$, $\beta$, $\gamma$, $\eta \in \lbrace x,y,z \rbrace$ denote the elements in cartesian coordinates, i.e. $r_{l\alpha}$
denotes component $\alpha$ of the vector $\textbf{r}_{l}$, which has the length $r_l=\left|\textbf{r}_{l}\right|$.

Our numerical values for the multipoles are given in Table~\ref{tab:2Multipoles} in Appendix~\ref{App:multipole_table}.  It is seen that the monopole ($q$) and dipole 
moments ($p_{\alpha}$) are essentially
zero, whereas the quadrupole moment ($Q_{\alpha \beta}$) is not.
However, as we will see below, the quadrupole moment alone is not sufficient to correctly describe the molecule's orientation to the substrate. 
We note in passing that our quadrupole moment for the planar configuration is very close to that obtained in Ref.~\cite{Sala2011}. One
also sees from Table~\ref{tab:2Multipoles} in Appendix~\ref{App:multipole_table} that the octupole ($O_{\alpha \beta \gamma}$) and the hexadecapole ($H_{\alpha \beta \gamma \eta}$) are also non-zero. 
We will come back to this point in the discussion of the molecule-substrate interaction. 

\subsubsection{Electrostatic pair interaction}\label{Sec:2B2_electrostat_pair}

To describe the electrostatic part of the molecule-molecule interaction we focus on the first non-vanishing multipole contribution, that is, the interaction stemming from the quadrupole
moments ($Q_{\alpha \beta}$). As seen from Table~\ref{tab:2Multipoles} in Appendix~\ref{App:multipole_table}, one has $Q_{x^, x^,}\approx Q_{y^, y^,}>0$, $Q_{z^, z^,}\approx -2\,Q_{x^, x^,}$ 
[in the eigensystem of the particle, see Fig.~\ref{fig:1_mol_model}(a)] for both, the planar and the twisted configuration. It therefore seems justified to approximate the full quadrupole-tensor
by that related to a \textit{linear} quadrupole. 
The latter is equivalent to three charges of magnitude $-q/2$, $q$, and $-q/2$ 
 that lie aligned, separated by equal distances $D$, on the $z^,$ eigenaxis of the original molecule [see Fig.~\ref{fig:1_mol_model}(a)]. The corresponding value of $Q$ is $Q=qD^2$.
%that lie aligned and equidistant on the $z^,$ eigenaxis of the original molecule
%The latter has the charge distribution ($-q/2$ \quad $+q$ \quad $-q/2$) and lies parallel to the $z^,$ eigenaxis of the original molecule
 
In the following we denote the direction of this linear quadrupole by the vector $\textbf{q}_i$. Note that $\textbf{q}_i$ lies \textit{perpendicular}
to the symmetry axis $\textbf{u}_i$ of the uniaxial ellipsoid introduced in Sec.~\ref{Sec:2A_non_electrostat}. In other words, the coarse-grained particle now becomes effectively biaxial.
Consequently, the orientation of $\textbf{q}_i$ in the space-fixed coordinate system is characterized by the three (Euler) angles, that is,

\begin{equation}
 \textbf{q}_i (\alpha, \vartheta, \varphi)=\begin{pmatrix} \cos\alpha \cos\vartheta \cos\varphi + \sin\alpha\sin\varphi\\
 \cos\alpha \cos\vartheta \sin\varphi - \sin\alpha\cos\varphi \\
 -\cos\alpha\sin\vartheta\end{pmatrix}\textrm{.}
\end{equation}

\begin{figure}
\includegraphics[bb=0 0 300 150]{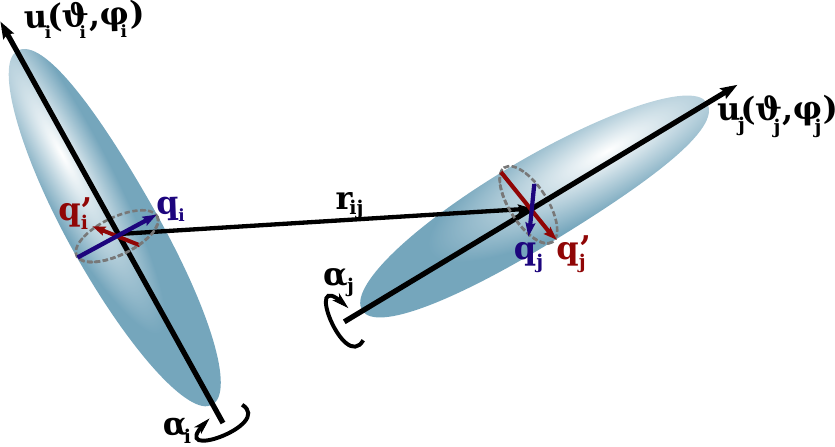}
\caption{(Color online) Exemplary configuration of two coarse-grained 6P molecules with the long axes $\hat{\textbf{u}}_i$ and $\hat{\textbf{u}}_j$ and the linear 
quadrupole moments oriented along \textbf{q}$_i$ and \textbf{q}$_j$. The quadrupole moments \textbf{q}$_i^{,}$ and \textbf{q}$_j^{,}$ 
denote the quadrupole moments that mimic the electrostatic molecule-substrate interaction [see Eq.~(\ref{eqn:Q_for_QS})].}
\label{fig:3_mol-mol_config}
\end{figure}

We are now in the position to write down the electrostatic interaction between two coarse-grained 6P molecules. In our model, this is the interaction between two linear 
quadrupoles \cite{Gray1984} given by

\begin{align}
\notag V_{QQ}(\textbf{q}_i, \textbf{q}_j, \textbf{r}_{ij})&=\frac{3}{4}\frac{1}{4 \pi \epsilon_0}\frac{Q^{2}}{\textrm{r}_{ij}^5}[1-5\cos\beta_i^2\\
\notag &-5\cos\beta_j^2-15 \cos\beta_i^2\cos\beta_j^2\\
 &+2(\cos\gamma_{ij}-5\cos\beta_i\cos\beta_j)^2]\textrm{,}\label{eq:V_QQ}
\end{align}
where $\textrm{r}_{ij}=\left| \textbf{r}_{ij}\right|$, $\cos\beta_i=\textbf{q}_i\cdot\hat{\textbf{r}}_{ij}$ (with $\hat{\textbf{r}}_{ij}=
\textbf{r}_{ij}/\textrm{r}_{ij}$), $\cos\beta_j=\textbf{q}_j\cdot\hat{\textbf{r}}_{ij}$ and $\cos\gamma_{ij}=\textbf{q}_i\textbf{q}_j$. The energetically most relevant 
configurations of two linear quadrupoles are discussed in Appendix~\ref{App:QQ_configs}. 
Furthermore, $Q=Q_{z^, z^,}$ (see Table~\ref{tab:2Multipoles} in Appendix~\ref{App:multipole_table}). Taken altogether, the total Hamiltonian for the molecule-molecule interaction is given by 

\begin{equation}
H_{\textrm{mol-mol}}=\sum_{i=1}^N \sum_{j \neq i} \left(V_{\textrm{GB}}(\hat{\textbf{u}}_i, \hat{\textbf{u}}_j, \textbf{r}_{ij}) + 
V_{\textrm{QQ}}(\textbf{q}_i, \textbf{q}_j, \textbf{r}_{ij})\right)\textrm{,}\label{H_mol_mol}
\end{equation}
where $V_{\textrm{GB}}$ and $V_{\textrm{QQ}}$ are defined in Eqs.~(\ref{eq:GB}) and (\ref{eq:V_QQ}), respectively. An exemplary configuration of two coarse-grained 6P 
molecules is shown in Fig.~\ref{fig:3_mol-mol_config}. Furthermore, numerical results for the various types of molecule-molecule interactions as a function of the distance are plotted in 
Fig.~\ref{fig:2_interactions}. The lowest-energy configuration corresponds to parallel oriented ellipsoids with a T-like orientation of the quadrupole moments.

\subsubsection{Electrostatic molecule-substrate interaction}\label{Sec:2B3_electr_field}

We now turn to the construction of an effective molecule-substrate interaction mimicking the influence of electrostatics. Our starting point are the molecular multipoles
introduced in Eqs.~(\ref{Eq:Monopole})-(\ref{Eq:Hexadec}). As we will see below, the quadrupole moment alone is not sufficient to correctly describe the molecule's
orientation to the substrate.

\begin{figure}
\includegraphics[bb=0 0 300 250]{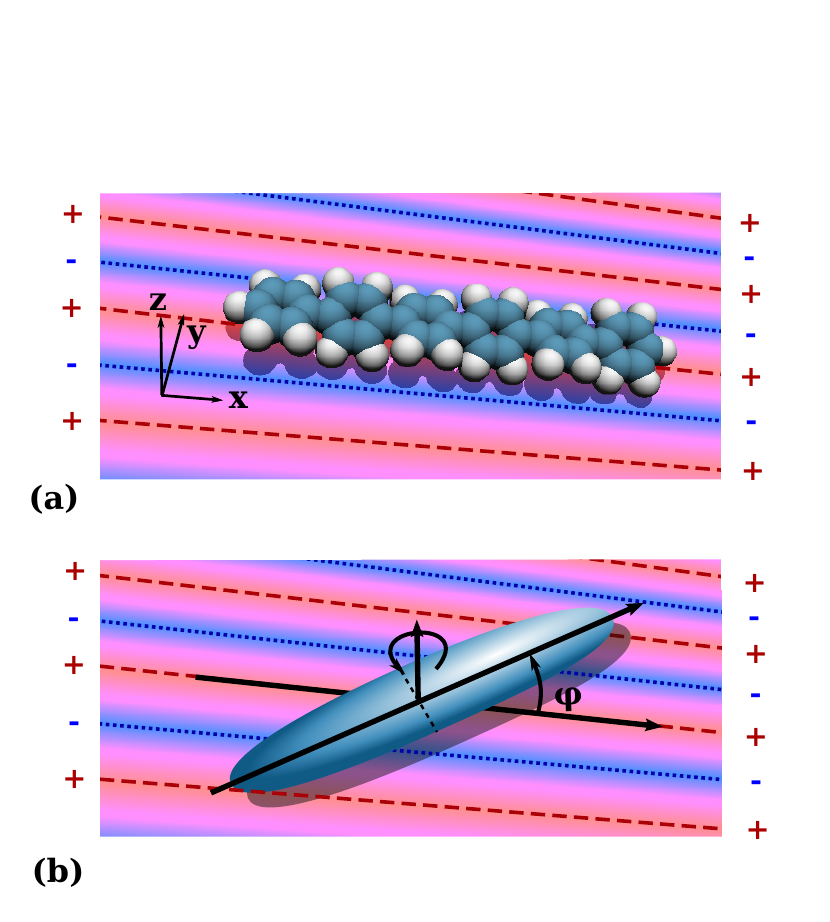}
\caption{(Color online) 6P molecule on the sketched ZnO(10-10) substrate (a) in its energetically most favored configuration and 
(b) in a coarse-grained representation, which illustrates the rotation discussed in Fig.~\ref{fig:5_multipoles}. The substrate
is characterized by line charges with alternating sign, as indicated by the red (dashed) and blue (dotted) lines for the positive and 
negative charges, respectively. The angle $\varphi$ describes the rotation with respect to the charge lines.}
\label{fig:4_mol_subs}
\end{figure}

In order to evaluate the importance of various moments, we consider the corresponding interaction energies $U_m(\textbf{r})$ [with $m$ referring to a specific multipole moment] 
in presence of an external (substrate) 
field $\tilde{\textbf{E}}(\textbf{r})$. The corresponding expressions for a molecule with center-of-mass position $\textbf{r}$ are given by

\begin{align}
 U_p(\textbf{r})&=- \sum_{\alpha} p_{\alpha} \tilde{E}_{\alpha} (\textbf{r})\textrm{,}\label{Eq:E_dipole}\\
 U_Q(\textbf{r})&=-\frac{1}{3} \sum_{\alpha,\beta} Q_{\alpha \beta} \frac{\partial\tilde{E}_\alpha }{\partial x_\beta} (\textbf{r})\textrm{,}\label{Eq:E_quadrupole}\\
 U_O(\textbf{r})&=-\frac{1}{15} \sum_{\alpha,\beta,\gamma} O_{\alpha \beta \gamma} \frac{\partial^2\tilde{E}_{\alpha} }{\partial x_{\beta}\partial x_{\gamma}} (\textbf{r})\textrm{,}\\
 U_H(\textbf{r})&=-\frac{1}{105} \sum_{\alpha,\beta,\gamma,\eta} H_{\alpha \beta \gamma \eta} \frac{\partial^3\tilde{E}_{\alpha} }{\partial x_{\beta}\partial x_{\gamma}\partial x_{\eta}} (\textbf{r})\textrm{.}\label{Eq:E_hexadec}
\end{align}
Here, $\partial/\partial x_{\alpha}$ stand for derivatives with respect to cartesian coordinates. In the present study, the field $\tilde{\textbf{E}}(\textbf{r})$ stems from the ZnO(10-10) substrate, which is characterized by a so-called 
``mixed'' termination: the substrate features alternating lines 
of Zn atoms and O atoms \cite{Blumstengel2010}. Since these have different effective charges, the ZnO(10-10) substrate effectively displays alternating, 
parallel rows of positive and negative charges. In the following we assume that these lines are oriented along the $x$-axis of the
coordinate system (see Fig.~\ref{fig:4_mol_subs}). According to Ref.~\cite{Sala2011}, the resulting electrostatic field can be approximated as 
\begin{equation}
 \tilde{\textbf{E}}(\textbf{r})=\begin{pmatrix} 0 \\A \exp(-k\textrm{z}) \cos(k\textrm{y}) \\ -A \exp(-k\textrm{z}) \sin(k\textrm{y}) \end{pmatrix}\textrm{,} \label{eq:field}
\end{equation}
where $A\approx97$~eV/(nm e) is the field strength and $k=2 \pi/0.519$~nm is the wave length of a substrate unit cell measured in $y$-direction 
(i.e., orthogonal to the charge lines on the substrate plane). With the field given in Eq.~(\ref{eq:field}) we can evaluate the electrostatic energy contributions given in 
Eqs.~(\ref{Eq:E_dipole})-(\ref{Eq:E_hexadec}) as functions of all molecular degrees of freedom, that is
the position \textbf{r}, and the angles $\varphi$, $\vartheta$ and $\alpha$. To investigate the importance of the multipoles, we choose
\textbf{r}, $\vartheta$ and $\alpha$ according to an energetic minimum~\cite{Sala2011}, 
that is, $x=0$~nm, $y=0.519\textrm{~nm}\cdot 3/4=0.389$~nm, $z=z_{\textrm{min}}$, and $\alpha=0$, $\vartheta=\pi/2$ (in-plane configuration). The quantity of interest is then
the rotational energy $E_{\textrm{electr}}(\varphi)$, where $\varphi$ is the in-plane angle relative to the $x$-direction.

\begin{center}
\begin{figure*}
\begin{minipage}{17cm}
\includegraphics[bb=0 0 500 150]{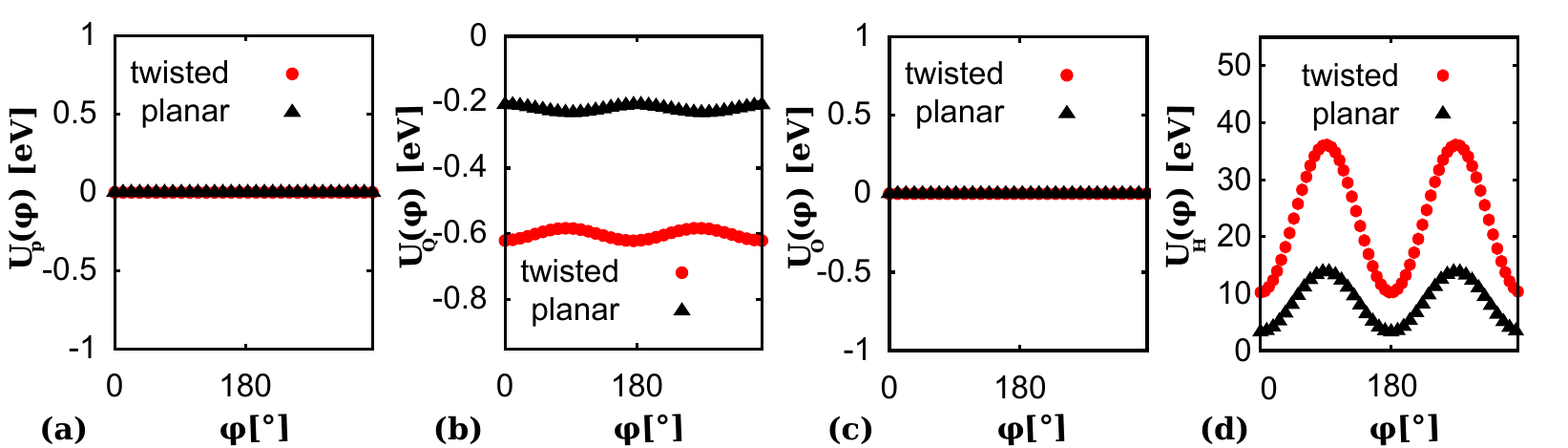}
\caption{(Color online) Contributions to the in-plane rotational energy from the molecular multipole moments in the electrostatic field of the ZnO(10-10) substrate [(a) dipole,
(b) quadrupole, (c) octupole, (d) hexadecapole].
In all sub-figures, $\varphi$ denotes the angle by which the molecule is rotated with respect to the charge lines. 
During this rotation the shortest axis of the molecule, $z^,$, forms the axis of rotation and thus remains perpendicular to the substrate, which is equivalent to saying 
that $z^,$ lies parallel to $z$ [see Fig.~\ref{fig:4_mol_subs}(b)]. The molecules considered are 6P in its planar configuration (planar) and a twisted configuration that the molecule assumes in vacuum 
(twisted) \cite{Palczynski2014, priv_Palczynski}.}
\label{fig:5_multipoles}
\end{minipage}
\end{figure*}
\end{center}

We determine the rotational energy
for the different multipole moments $E_{\textrm{electr}}(\varphi)=U_p(\varphi)\textrm{, }U_Q(\varphi)\textrm{, }U_O(\varphi)\textrm{, }U_H(\varphi)$ individually.
To this end we rotate every atomic position $\textbf{r}_{l}$ by $\varphi$ around the center of the molecule, evaluate the multipole moments 
via Eqs.~(\ref{Eq:Dipole})-(\ref{Eq:Hexadec})
and finally determine the energy of the respective multipole in the field given in Eq.~(\ref{eq:field}) using Eqs.~(\ref{Eq:E_dipole})-(\ref{Eq:E_hexadec}).
The resulting energy functions are depicted in Fig.~\ref{fig:5_multipoles}(a)-(d), respectively.

It is seen that the first (dipole) and the third (octupole) moment 
do not contribute at all to the in-plane rotational energy [see Figs.~\ref{fig:5_multipoles}(a) and (c)]. The second (quadrupole) moment does contribute to the rotational energy,
however, it contributes with 
an approximately constant value [see Fig.~\ref{fig:5_multipoles}(b)]. The absence of a clear minimum means that the quadrupole does \textit{not} favor the 
alignment of the molecule with the charge lines on the substrate. 
The first significant contribution to the molecular orientation arises through the interaction of the fourth (hexadecapole) moment with the substrate field 
[see Fig.~\ref{fig:5_multipoles}(d)]. 
This rotational energy is minimal for a molecule that is aligned with the charge lines of the substrate for both, the planar and the twisted configuration (even though the
actual values do depend on the configuration).
We therefore conclude that the hexadecapole moment is responsible for the alignment observed by Della Sala \textit{et al.} \cite{Sala2011}. 

However, from a computational point of view, the treatment of the hexadecapole-field interaction for \textit{many} molecules implies a very large effort. We 
therefore \textit{mimic} the effect of this interaction by assigning to each molecule an additional, fictitious quadrupole, which is linear in character and is oriented
\textit{perpendicular} to the original one. We call this quadrupole $\textbf{q}^,_i$. The value of the corresponding moment is discussed in Sec.~\ref{Sec:2B4_parametrization}.
The \textit{effective} electrostatic part of the molecule-substrate interaction is then given by
\begin{align}
 \notag V_{\textrm{QS}}(\textbf{q}^,_i &, \textbf{r}_{i})=-\frac{1}{3} \\
 &\times\sum_{\alpha \beta}\left(\sum_{\gamma \eta} R_{\alpha \gamma}(\textbf{q}^,_i) Q_{\gamma \eta}^, R_{\beta \eta}(\textbf{q}^,_i)\right) \left. \frac{\partial \tilde{E}_{\alpha}}{\partial x_{\beta}}\right|_{\textbf{r}_{i}}\textrm{.}\label{eq:QS}
\end{align}
Here $Q_{\alpha \beta}^,$ denotes the \textit{effective} quadrupole tensor of the molecule in the molecule's eigensystem (see Sec.~\ref{Sec:2B4_parametrization}), and
 $R=R(\varphi,\vartheta,\alpha)$ is a conventional rotation tensor in euclidean space involving the Euler angles ($\varphi,\vartheta,\alpha$).

Using Eq.~(\ref{eq:QS}) for the electrostatic molecule-substrate interaction $V_{\textrm{QS}}(\textbf{q}^,_i, \textbf{r}_{i}))$ and Eq.~(\ref{eq:V_LJ})
for the non-electrostatic interaction $V_{\textrm{LJ}}(\hat{\textbf{u}}_i)$, we obtain the total Hamiltonian for the molecule-substrate interaction,
\begin{equation}
H_{\textrm{mol-subs}}=\sum_{i=1}^N\big( V_{\textrm{LJ}}(\hat{\textbf{u}}_i) + V_{\textrm{QS}}(\textbf{q}^,_i, \textbf{r}_{i}) \big)\textrm{.}\label{eq:mol_subs}
\end{equation}

\subsubsection{Parametrization of the quadrupole-field interaction}\label{Sec:2B4_parametrization}

\begin{figure}
\includegraphics[bb=0 0 400 400]{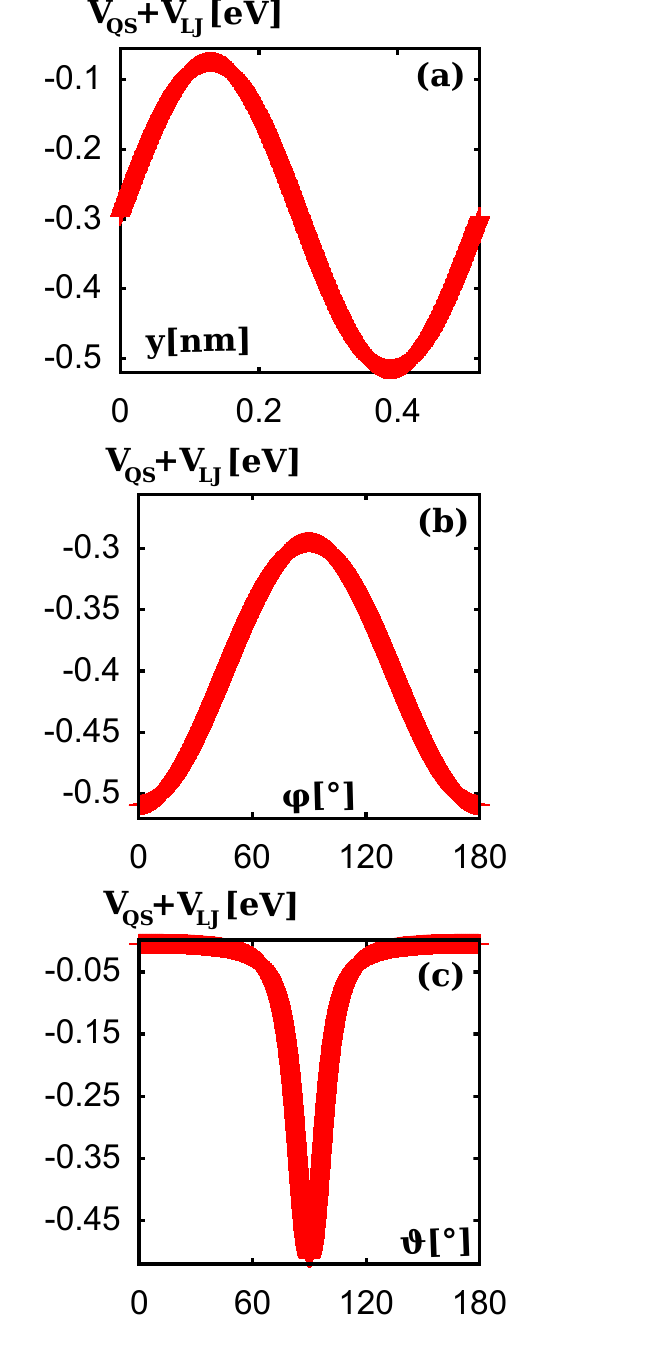}
\caption{(Color online) Interaction potential of a single 6P molecule with the Zn0(10-10) substrate according to Eq.~(\ref{eq:mol_subs}) at $z=z_{\textrm{min}}$. 
Part (a) shows the potential as a function of the $y$-coordinate. 
At $y=0.129$~nm the molecule lies above a negative charge (energetic maximum), while at $y=0.389$~nm it lies above a positive charge (energetic minimum). Parts (b) and (c) show the potential as a function of the 
angles $\varphi$ and $\vartheta$.}
\label{fig:6_rot_pot}
\end{figure}

To adjust the magnitude of the fictitious linear quadrupole introduced in Eq.~(\ref{eq:QS}) we make use of the value for the rotational energy barrier $\Delta E_{\textrm{r}}$ given in 
Ref.~\cite{Sala2011}. For a molecule which lies flat on the substrate and is rotated around the $z$-axis, the authors in Ref.~\cite{Sala2011} report a value
of $\Delta E_{\textrm{r}}=220$~meV. 
In our case the corresponding barrier is given by

\begin{align}
\notag \Delta E_{\textrm{r}}&=\Bigg(V_{\textrm{QS}}(\varphi=0)\\
&-V_{\textrm{QS}}(\varphi=90^{\circ})\Bigg)_{\vartheta=\pi/2, \alpha=0, y=0.389\textrm{~nm}}\textrm{,}
\end{align}
where the electrostatic molecule-substrate interaction $V_{\textrm{QS}}(\textbf{q}^,_i, \textbf{r}_{i}))$ is defined in Eq.~(\ref{eq:QS}).

It turns out that we can reproduce the literature value for $\Delta E_{\textrm{r}}$, as well as the functional form of the molecule-substrate potential with respect
to $y$ and $\varphi$, by using a linear quadrupole $\textbf{q}^,_i$ corresponding to a quadrupole tensor of the form

\begin{equation}
 Q^,=\begin{pmatrix} -0.013 & 0 & 0 \\0 &0.026 &0 \\ 0 & 0& -0.013 \end{pmatrix}\textrm{~e nm}^2\textrm{.}\label{eqn:Q_for_QS}
\end{equation}

The quadrupole element $Q^,_{\textrm{x}^,\textrm{x}^,}$ describes the quadrupole strength parallel to the longest axis of the molecule and $Q^,_{\textrm{z}^,\textrm{z}^,}$ is 
the quadrupole strength parallel to the shortest axis
(see Fig.~\ref{fig:1_mol_model}). The linear quadrupole $\textbf{q}^,_i$ defined by Eq.~(\ref{eqn:Q_for_QS}) is orthogonal to both the long axis of the molecule $\textbf{u}_i$ and 
the orientation of the quadrupole direction $\textbf{q}_i$ used for the electrostatic molecule-molecule interaction [as is depicted in Fig.~\ref{fig:3_mol-mol_config}]. The latter is oriented parallel to the $z^,$-axis of the molecule (see Sec.~\ref{Sec:2A_non_electrostat}). 
Therefore, its orientation is not influenced if the molecule lies on the substrate and is rotated around its $z^,$-axis by $\varphi$. 
This explains why the linear quadrupole $\textbf{q}_i$ cannot account for 
the orientation of the molecule within the plane.

The full molecular-substrate potential [see Eq.~(\ref{eq:mol_subs})] as function of $y$, $\vartheta$ and $\varphi$ using the parametrization given in Eq.~(\ref{eqn:Q_for_QS}) is plotted in 
Fig.~\ref{fig:6_rot_pot}. The translational energy ``landscape'' depicted in Fig.~\ref{fig:6_rot_pot}(a) agrees on a qualitative level with 
the corresponding landscape 
for
2P on ZnO(10-10) found by Della Sala \textit{et al.} \cite{Sala2011} (this study does not include corresponding results for 6P). Moreover, qualitative and 
\textit{quantitative} agreement is achieved for the rotational energy plotted in Fig.~\ref{fig:6_rot_pot}(b). In particular,
the energy barrier height $\Delta E_{\textrm{r}}\approx 220$~meV fully reproduces the value given in Ref.~\cite{Sala2011}.

\section{Ordering behavior of an ensemble of 6P molecules}\label{Sec:3Ordering}

In this section we employ the model Hamiltonian given in Eqs.~(\ref{H_mol_mol}) and (\ref{eq:mol_subs}) to get first insights into the corresponding many-particle
behavior. To this end we use
 MC simulations of a system with fixed particle x- and y-positions,
 but continuous 3D particle orientations. 
%To this end we use MC)simulations of a system with fixed particle positions, but continuous 3D particle orientations. 
Thus,
our focus in this study is to understand the orientational ordering at different densities. Some methodological details are presented in the subsequent sections~\ref{Sec:3A_simulation} and
\ref{sec:3B_evaluation}. In the remainder of section~\ref{Sec:3Ordering} we then discuss the numerical results.

\subsection{Monte-Carlo simulation}\label{Sec:3A_simulation}

\begin{figure}
    \includegraphics[bb=0 0 300 100]{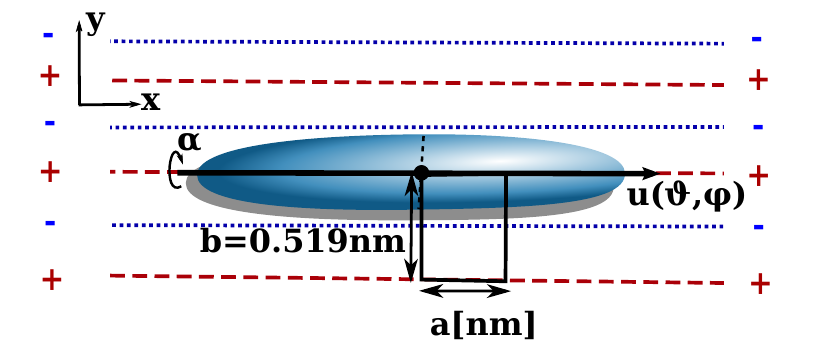}
\caption{(Color online) Sketched configuration of a molecule on the substrate lattice. The center of mass sits at a position on the positive charge lines [see Fig.~\ref{fig:4_mol_subs}(a)]. 
Two positive (negative) charge lines have a distance $b$, while the lattice constant $a$ in $x$ direction is given by $a=0.329$~nm \cite{Parker1998}. 
 The molecule has continuous 3D orientation described by the Euler angles 
$\varphi$, $\vartheta$ and $\alpha$.}
\label{fig:7_surface_unit}
\end{figure}

Our simulations are based on a two-dimensional lattice consisting of either square unit cells (lattice constant $a$) or tetragonal unit cells characterized by
lattice constants $a\neq b$. The tetragonal cell is inspired by the unit cell of the real ZnO(10-10) substrate where $a=0.329$~nm and $b=0.519$~nm \cite{Parker1998}.
A corresponding sketch is given in Fig.~\ref{fig:7_surface_unit}. For both lattice types, we set the number of particles per unit cell to one; moreover,
this molecule's center of mass is fixed in lateral directions. Specifically, we set $x=n\cdot a$ and $y=m\cdot b+3\, b/4$ 
, where $n$, $m$ are integers (recall that $y=3\, b/4=0.389$~nm is the position of the energetic minimum in $y$ direction, see Fig.~\ref{fig:6_rot_pot}). Thus,
the main degrees of freedom are the three Euler angles $\vartheta$, $\varphi$, $\alpha$ introduced in Fig.~\ref{fig:1_mol_model}(c). Note that the rotational freedom of the particles implies that their center of mass can have different distances from the surface, since
$z_i=z_i(\textbf{u}_i)$ [see Eq.~(\ref{eqn:z_vartheta})]. As a consequence, our simulations allow for both, lying and standing configurations. \\

The acceptance probability for each rotational move of molecule $i$ (involving all three Euler angles) is given by the conventional Metropolis scheme, that is 
\begin{equation}
 p_{i}^{R}= \min \left\{ \exp\left(\frac{H_{\textrm{initial}}(i) -H_{\textrm{final}}(i)}{kT}\right) ,1\right\}\textrm{,}
\end{equation}
where the Hamiltonians $H$ for the initial and the final configuration are determined using Eq.~(\ref{eqn:hamiltonian}), (\ref{H_mol_mol}) and 
(\ref{eq:mol_subs}). For computational efficiency, we restrict the 
range of the molecule-molecule interaction to center-of-mass distances $r_{ij}<1.2\,l=3.25$~nm. 
The temperature $T$ is chosen as $T=300$~K unless specified otherwise.  In fact, the only system we have studied at 
temperatures below 300K are systems without molecule-molecule interaction, where frustration effects are absent per definition. 
At 300K, we have checked equilibration by testing different (randomly oriented) initial conditions and monitoring acceptance 
rates for rotational moves. These rotational moves have been carried out as follows:\\
The angles $\alpha$ and $\varphi$ are randomly selected from a normal distribution $\in \left[0,360^{\circ}\right]$,
while $\vartheta$ is drawn from the distribution $\textrm{arccos}(r-1)$, with $r$ being a random number chosen uniformly from the interval $\left[0,1\right]$.\\
Our aim is to understand the role of the individual contributions to the Hamiltonian for the overall orientational ordering behavior. To this end, we perform 
MC simulations for different subsets of the interactions determined in Eqs.~(\ref{H_mol_mol}) and (\ref{eq:mol_subs}).

First, for systems in which no molecule-substrate interactions are considered, we use square unit cells and vary the density by varying the lattice constant $a$.
Second, for systems with electrostatic field, we fix the lattice constant in $y$ direction to the distance of the substrate charge lines, i.e., $b=0.519$~nm. 
To vary the density, we then only vary the lattice constant $a$ in 
$x$-direction (see Fig.~\ref{fig:7_surface_unit}). 
Note that the molecule's length of $l=2.79$~nm (see Table~\ref{tab:1GBParam}) 
is much larger than $b=0.519$~nm. Thus, even for large values of $a$, the rotational motion of molecules in $y$ direction is strongly restricted.

The entire simulation box consists of at least 1000 unit cells for the square lattice and up to 2000 unit cells for 
simulations on the tetragonal lattice, in order to maintain reasonable statistics despite the broken lateral symmetry. 
We employ
periodic boundary conditions in the $x$- and $y$-directions. We initialize the lattice with randomly oriented molecules and then equilibrate the systems for at 
least $10^5$ MC steps, followed by production runs over another $10^5$ MC steps. 

\subsection{Measures of evaluation}\label{sec:3B_evaluation}

In order to analyze the orientational ordering of the system, we use the conventional, traceless second-rank tensor defined in Ref.~\cite{Tildesley2009, Strehober2013},
\begin{equation}
 A_{\alpha\beta}=\frac{1}{N}\sum_{i=1}^{N}\left\langle \textrm{u}_{i,\alpha} \cdot \textrm{u}_{i,\beta} -\frac{1}{3} \delta_{\alpha \beta} \textrm{Tr}(\textbf{u}_i \otimes \textbf{u}_i)\right\rangle \textrm{,}\label{eqn:S}
\end{equation}
where $\otimes$ stands for a dyad product, Tr is the trace and $\langle \dots \rangle$ denotes an average over an ensemble of configurations.

To determine the overall degree of (nematic) order, we consider the eigenvalues and eigenvectors of the tensor $A_{\alpha\beta}$. In a system with perfect uniaxial order, the set of
eigenvalues is of the form $\{ \mu_k \}=\{ -1/3, -1/3, 2/3 \}$. The eigenvector associated to the eigenvalue mu with the largest absolute value (i.e. $\left|\mu\right|\geq\left|\mu_k\right|$ for all $k$)) 
is the director of the system. Further, the largest absolute eigenvalue of $A_{\alpha\beta}$ is also directly proportional to the conventional Maier-Saupe order-parameter $S=3 \mu/2$. 
For a perfectly uniaxal system it follows that $S = 1$. In our system, large values of $S$ typically occur at large densities, where the molecules form a standing uniaxially ordered phase.

Another important situation occurs when the molecules' orientations are restricted to (arbitrary) directions within the plane, say, the $x$-$z$-plane. Then the eigenvalues of $A_{\alpha\beta}$ 
are of the form $\{ \mu_k \}=\{ 1/6, 1/6, -1/3 \}$, yielding $S = -1/2$ [56,M,N]. 
Here, we observe such negative values of $S$ on tetragonal lattices. Finally, $S = 0$ represents a completely disordered system.

In order to further characterize the orientation of the molecules, we determine various angular distributions involving the Euler angles $\varphi_i$, $\vartheta_i$ and $\alpha_i$. 
 Specifically we define the probability that a particle has an angle $\beta=\varphi$, $\alpha$, or $\beta=\varphi+\alpha$ as 
\begin{equation}
 P(\beta)=\left\langle \frac{n_{\left[\beta-\Delta\beta/2,\beta+\Delta\beta/2\right]}}{N\cdot \Delta \beta}\right\rangle\textrm{,} \label{eqn:P_angle}
\end{equation}
where $n_{\left[\beta-\Delta\beta/2,\beta+\Delta\beta/2\right]}$ is the number of particles in a tolerance interval defined by 
$\Delta \beta =6^{\circ}$. In the absence of orientational ordering these distributions are constant. Also note that $\alpha, \varphi, \alpha+\varphi$ take values between $0$ and $180^{\circ}$. 
To describe the distribution of the azimuthal angle $\vartheta$ we consider the quantity $P(\cos \vartheta)$ defined in accordance with Eq.~(\ref{eqn:P_angle}), but with $\Delta \beta =0.02$. 
The advantage of considering $P(\cos \vartheta)$ rather than $P(\vartheta)$ is that $P(\cos \vartheta)$ is constant in an isotropic phase, contrary to $P(\vartheta)$.

Further, we study the height distribution function $P(z)$. he height distribution function is defined in analogy to the angular distribution function $P(\beta)$ [see Eq.~(\ref{eqn:P_angle})], 
where the height interval used is $\Delta z=0.02$~nm. Note, however, that in our system $P(z)$ and $P(\cos \vartheta)$ are intimately related through Eq.~(\ref{eqn:z_vartheta}).

\subsection{Numerical results}\label{Sec:3C_results}

\subsubsection{Impact of molecule-molecule interactions on a quadratic lattice}\label{Sec:3C1_square}

\begin{center}
\begin{figure*}
\begin{minipage}{17cm}
\includegraphics[bb=0 0 500 200]{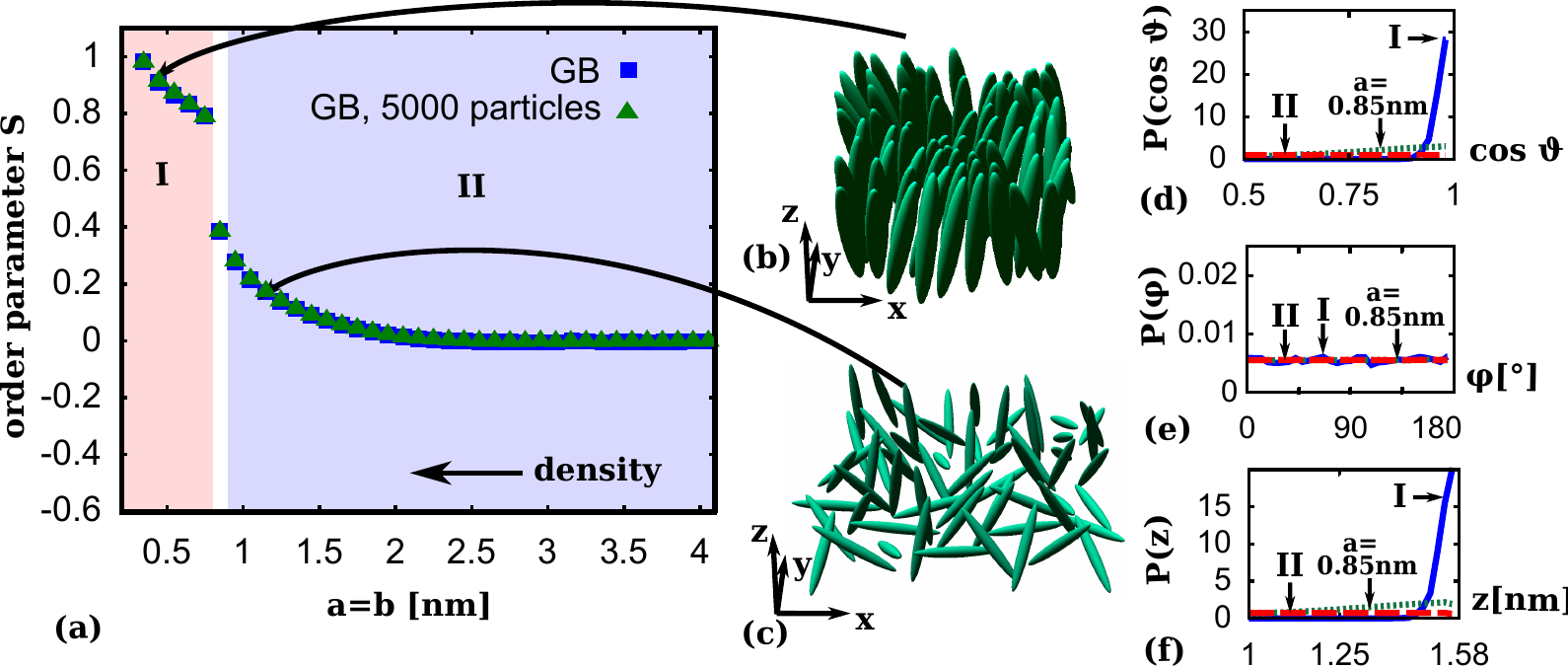}
\caption{(Color online) Ordering behavior of molecules interacting exclusively through GB interactions (with axis ratio 1:8.3, corresponding to 6P).
The $x$ and $y$ positions of the centers of mass are positioned on a square lattice with lattice constant $a=b$. Part (a) displays the nematic order parameter for different values of $a$ in a systems with 1000 
(blue squares) and 5000 particles (green triangles), respectively.  
The two observed orientational states are denoted by roman numbers: I stands for standing (upright) uniaxial order, II stands for 3D (isotropic) disorder.
Parts (b)-(c) show snapshots of a subset of the molecules at $a=0.4$~nm (state I) and $a=1.0$~nm (state II), respectively.
The two states are further characterized in parts (d)-(f), in which we present the angular distributions for 
$\cos\vartheta$ and $\varphi$ at $a=0.4$~nm (state I) and $a=4.0$~nm (state II), as well as the height distribution $P(z)$. Parts (d)-(f) also contain the angular distributions 
for a lattice constant $a=0.85$~nm, which corresponds to the last point of phase II before the transition to phase I.}
\label{fig:8_square_GB}
\end{minipage}
\end{figure*}
\end{center}

As a starting point, we focus on the orientational ordering of molecules on a quadratic lattice without a substrate pattern. 
The $z$-coordinates of the centers of mass are restricted to the intervall [$z_{\mathrm{min}},z_{\mathrm{min}}-d/2+l/2$] where $l/2$ is half of one
molecule's length and $d/2$ is half of one
molecule's diameter. Thus, the angular coordinate $\vartheta_i$ of each molecules can assume all possible values.
In Fig.~\ref{fig:8_square_GB}(a) we present MC results for the order parameter $S$ in a system of molecules 
that interact exclusively through GB interactions for different values of the lattice constant $a$. As $a$ increases, the density of the molecules decreases. 
Inspecting the function $S(a)$ in Fig.~\ref{fig:8_square_GB}(a) one can identify two different regions.
In the first region (state I) occurring at small values of $a$ (i.e., large densities), the system displays large, positive values of $S$, indicating uniaxial ordering.
The director of this ordering points along the $z$-direction, i.e. the molecules stand upright, as illustrated by the snapshot in Fig.~\ref{fig:8_square_GB}(b) and by
the peak of the distribution function $P(\cos \vartheta)$ at
$\cos \vartheta=1$ in Fig.~\ref{fig:8_square_GB}(d). As expected, the corresponding distribution $P(\varphi)$ in Fig.~\ref{fig:8_square_GB}(e) is essentially flat.
This uniaxial upright ordering is indeed expected in view of the fact that the lattice
constant is significantly smaller than the molecule's length. We also note that uniaxial ordering is a generic feature of dense systems of elongated particles,
even when the interactions are purely repulsive \cite{Bolhuis1997}.

Increasing the lattice constant $a$ towards larger values there appears a second region, where $S$ takes initially small, yet non-zero values and eventually approaches zero upon further increase of $a$. In this region (II), the molecules have full rotational freedom as reflected by the 
snapshot in Fig.~\ref{fig:8_square_GB}(c) and by the nearly flat distributions of $\cos\vartheta$ and $\varphi$ in 
Fig.~\ref{fig:8_square_GB}(d), (e). These features correspond to a 3D disordered phase. 
Note that at lattice constants that are within region II, but close to the transition to region I (e.g. at $a=0.85$~nm) the values of S become relatively large ($S\approx0.4$). 
We find that these values are uninfluenced through system size [see Fig.~\ref{fig:8_square_GB}(a)]. Test simulations for lower temperatures show that the value 
of $S$ before the transition decreases to $S\approx0.05$. At $300$K and $a=0.85$~nm, the system displays a slight collective (i.e. system-averaged) ordering in $\vartheta$ and no collective order in $\varphi$, as is depicted in Fig.~\ref{fig:8_square_GB}(d) and (e). 
Thus, we find a finite degree of collective ordering in the ``isotropic'' phase very close to the boundary to phase I. 

Finally, in Fig.~\ref{fig:8_square_GB}(f), we show the distribution $P(z)$. Phase II is characterized by a homogeneous distribution of
$z$-values within the accessible interval. In contrast, we find a clear peak at
$z\approx1.57$~nm for phase I. This peak corresponds to an upright ordering, as is discussed in the text under Eq.~(\ref{eqn:z_vartheta}).

\begin{figure}
   \includegraphics[bb=0 0 300 200]{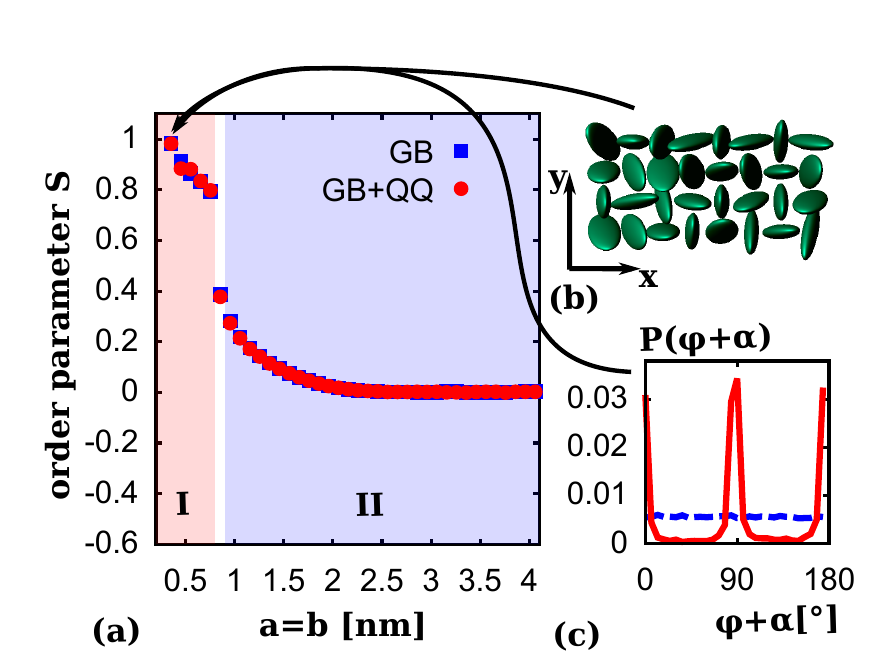}
\caption{(Color online) Ordering behavior induced by the full molecule-molecule interactions on a square lattice. In part (a), red dots
indicate the values of $S$ for GB plus quadrupolar (QQ) interactions; the corresponding values for a pure GB system are included as a reference (blue squares).
The snapshot in part (b) pertains to $a=0.4$~nm, where the molecules display a standing uniaxial order of their long axes combined with a T-shaped ordering
of their quadrupoles. In order to better visualize this situation, the particles are shown flattened along their $z^,$ axis. 
Part (c) shows the corresponding distribution $P(\varphi+\alpha)$ for pure GB (blue dashed) and GB plus quadrupolar interactions (red). 
For standing molecules ($\vartheta_i=0$), the angle $\varphi+\alpha$ denotes the rotation of the molecule around its long ($x^,$)-axis.}
\label{fig:9_square_QQ}
\end{figure}

We now consider the impact of the additonal electrostatic interactions induced by the quadrupole moments $\textbf{q}_i$ oriented perpendicular to the molecule's 
long axis $\textbf{u}_i$ (see Fig.~\ref{fig:3_mol-mol_config}). Corresponding results for $S$ as function of $a$ are plotted in Fig.~\ref{fig:9_square_QQ}(a), where we have included the data from 
Fig.~\ref{fig:8_square_GB}(a) for the pure GB system as a reference.
 It is seen that the quadrupolar interactions (QQ) do not significantly change the magnitude 
of $S$, indicating that the general phase behavior remains unchanged. We note in this context that the definition of $S$ involves the directions $\textbf{u}_i$ alone. However, one marked difference occurs when we analyze 
the local structure within the upright uniaxially ordered phase: In the system with quadrupolar interactions, neighboring molecules tend to order into T-shaped configurations with respect to the directions 
of their quadrupole moments, see Fig.~\ref{fig:9_square_QQ}(b). This T-like ordering is also reflected by the distribution function $P(\varphi+\alpha)$ plotted in
Fig.~\ref{fig:9_square_QQ}(c). It is seen that $P(\varphi+\alpha)$ has two pronounced (and equally high) peaks at $\pi/2$ and $\pi$, indicating that these
are the favored orientations of the $\textbf{q}_i$.
In fact, the resulting structure somewhat resembles that in the ``herringbone phase'', which has been observed in real 6P systems \cite{Socci1993, Resel2008}. 
In these real systems, the angle between neighboring molecules is typically less than ninty degrees \cite{Al-Shamery2009}. In our system,
the angle of about $90^{\circ}$ between the quadrupoles of neighboring molecules can be explained by the fact that
the T- configuration is indeed the one with the lowest pair energy (see Appendix~\ref{App:QQ_configs}), and that this configuration
is compatible with the square lattice. In the sense that the square lattice structure {\em stabilizes} the T-like alignment of the quadrupoles. 
If the positions of molecules were allowed to freely vary, we would rather expect a herringbone orientational structure. 

\subsubsection{Tetragonal lattice}\label{Sec:3C2_tetragonal}
\begin{figure}
    \includegraphics[bb=0 0 300 150]{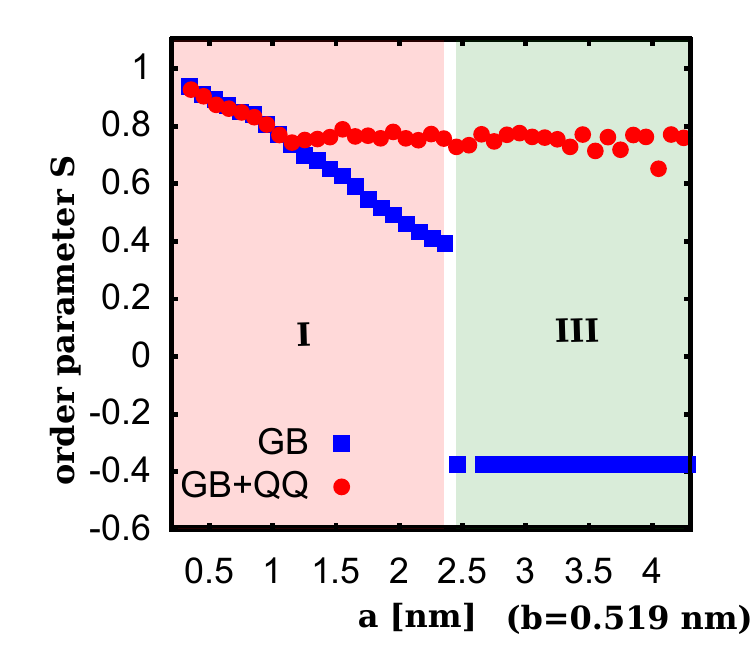}
\caption{(Color online) 
Nematic order parameter of 6P molecules interacting through GB (blue squares) as well as full molecule-molecule interactions 
(red dots) on a tetragonal lattice with varying lattice constant $a$. The different ordering states are denoted through roman numerals: 
I upright uniaxial order, III disorder within the $x$-$z$-plane.
These states are further analyzed in Fig.~\ref{fig:11_tetra_GB_part2}.}
\label{fig:10_tetra_GB}
\end{figure}

\begin{figure}
    \includegraphics[bb=0 0 300 300]{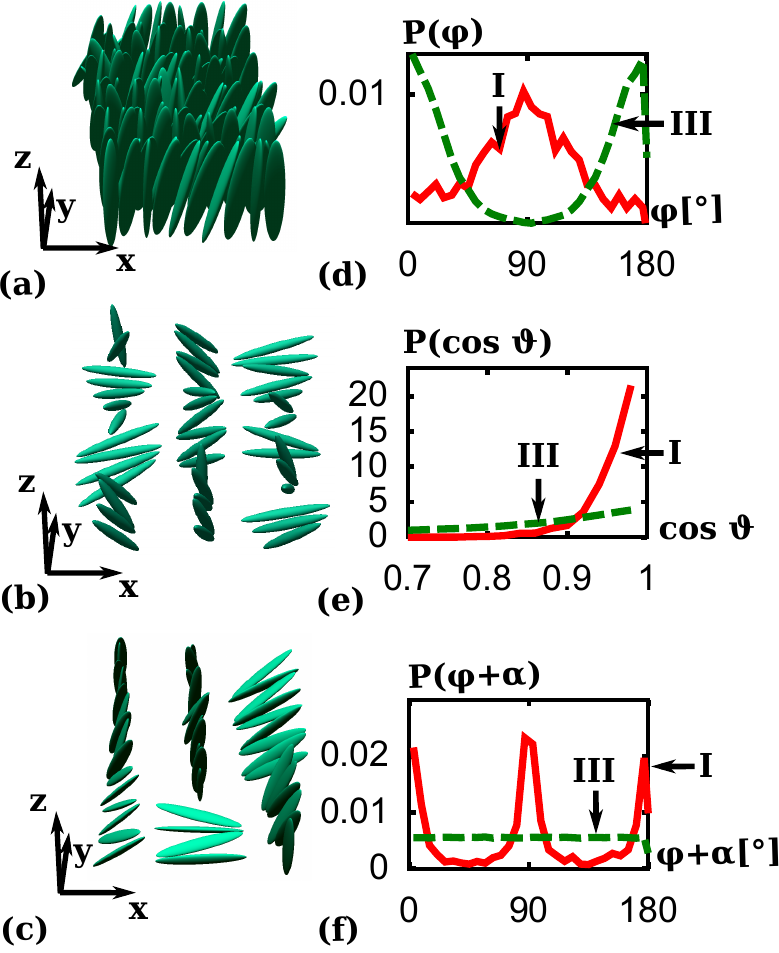}
\caption{(Color online) 
Orientational states of 6P molecules with different molecule-molecule (but no molecule-substrate) interactions. 
Parts (a) and (b) contain typical snapshots in region I ($a=0.4$~nm, GB+QQ) and III ($a=2.5$~nm, GB only), respectively. 
Corresponding angular distributions are plotted in parts (d)-(f). For additional insight, part (c) contains a snapshot 
for GB+QQ interacting molecules at lower density ($a=2.5$~nm).}
\label{fig:11_tetra_GB_part2}
\end{figure}

We now turn to the ordering behavior on a tetragonal lattice. As argued in section~\ref{Sec:3A_simulation}, tetragonal unit cells are characteristic of real ZnO(10-10) surfaces
with substrate pattern. As a background for this latter case, which will be discussed in section~\ref{Sec:3C3_substrate}, we investigate in the present subsection the impact
of the tetragonal lattice for (full) molecule-molecule interactions alone. 
Recall that on the tetragonal lattice, we vary the density by varying solely the lattice constant $a$, while the second lattice constant is kept fixed at $b=0.519$~nm.
Figure~\ref{fig:10_tetra_GB} plots the order parameter $S$ as a function of $a$, where we have
included data for both, systems with GB plus quadrupole interactions and pure GB systems. Corresponding snapshots and orientational distribution functions
are shown in Fig.~\ref{fig:11_tetra_GB_part2}.
Comparing the functions $S(a)$ in Fig.~\ref{fig:10_tetra_GB}
with those plotted in Fig.~\ref{fig:9_square_QQ} (quadratic lattice)
we see that the lattice type has indeed a profound impact on the overall behavior. This holds particularly for larger values of $a$,
where, moreover, pronounced differences between the GB system with quadrupolar interactions and the pure GB fluid occur.

To start with, at the lowest $a$ considered ($a=0.2$~nm), both the pure GB and the fully interacting system (GB+QQ) on the tetragonal lattice display an upright uniaxial phase (I), as in the case 
of the square lattice. Also, the quadrupole interactions induce a preference of T-like configurations for neighboring particles,
as illustrated by the snapshot in Fig.~\ref{fig:11_tetra_GB_part2}(a) and the angular distribution $P(\alpha+\varphi)$ in Fig.~\ref{fig:11_tetra_GB_part2}(f). 
This herringbone-like ordering is most pronounced at $a\approx 0.519$~nm, where the lattice is approximately 
quadratic.

Upon increasing $a$, the order parameter first somewhat decreases for both type of systems. However, beyond a lattice constant of $a\approx 1.2$~nm dramatic differences
between the interaction models (and between the lattice types) occur. For molecules without electrostatic interactions (pure GB), 
the (upright) uniaxial order first remains up to $a\approx 2.4$~nm. 
Crossing this value, the tetragonal unit cell induces a transition into a globally disordered state (III). 
Specifically, the overall order parameter $S\approx-0.4$, which is indicative for molecules restricted orientationally to a plane, 
as was previously discussed in section~\ref{sec:3B_evaluation}. In the present case, this plane is the $x$-$z$ plane [see snapshot in Fig.~\ref{fig:11_tetra_GB_part2}(b), 
as well as the angular distributions $P(\phi)$ and $P(\cos\vartheta)$ in Fig.~\ref{fig:11_tetra_GB_part2}(e) and (f), respectively]. 
We interpret the restriction to the $x$-$z$ plane from the fact that, in our simulations, the lattice constant in the $y$-direction is fixed to a rather small value 
($b=0.519$~nm). This precludes the molecules to fully explore the orientational space (or even align) in $y$-direction. However, despite the overall disorder in 
the $x$-$z$ plane, one observes finite domains characterized by local alignment, as expected in a strongly coupled system. 

Including now the quadrupolar intermolecular interactions, the system behaves in a completly different way 
(see the data labelled GB+QQ in Fig.~\ref{fig:10_tetra_GB}): It does not leave the upright uniaxial phase even for very large values of $a$ 
[see snapshot in Fig.~\ref{fig:11_tetra_GB_part2}(c)].
At first sight, this uniaxial ordering seems somewhat surprising due to low density considered and the fact, that the pure GB system eventually forms
a disordered state. To understand the impact of the quadrupolar interactions in this regime we recall, first, that the nearest-neighbor distance in 
$y$-direction (i.e., the lattice constant)
is small ($b=0.519$~nm) even at low densities due to our way to vary the density. Thus, the quadrupoles feel each other even at large values of $a$. 
Second, the (point) quadrupoles sit in the molecules center of mass. 
By forming an upright uniaxial phase, the molecules can thus reduce their electrostatic energy, while
keeping still some orientational freedom (note that for standing molecules, fluctuations of the angle $\vartheta$
induce only small changes of the height, and thus, of the interacting energy, contrary to the situation for lying molecules).
Moreover, within this standing uniaxial order, the quadrupole-quadrupole interactions induce a T-like ordering, which is 
visualized in Fig.~\ref{fig:11_tetra_GB_part2}(c).

\subsubsection{Impact of the electrostatic substrate pattern}\label{Sec:3C3_substrate}
So far, we have been focusing on the role of the various molecule-molecule interactions on the overall
ordering behavior, while the substrate has been considered just as a confining medium. This seems appropriate 
if one considers, e.g., 6P at an oxygen-terminated ZnO(000-1) surface.

\begin{figure}
    \includegraphics[bb=0 0 300 250]{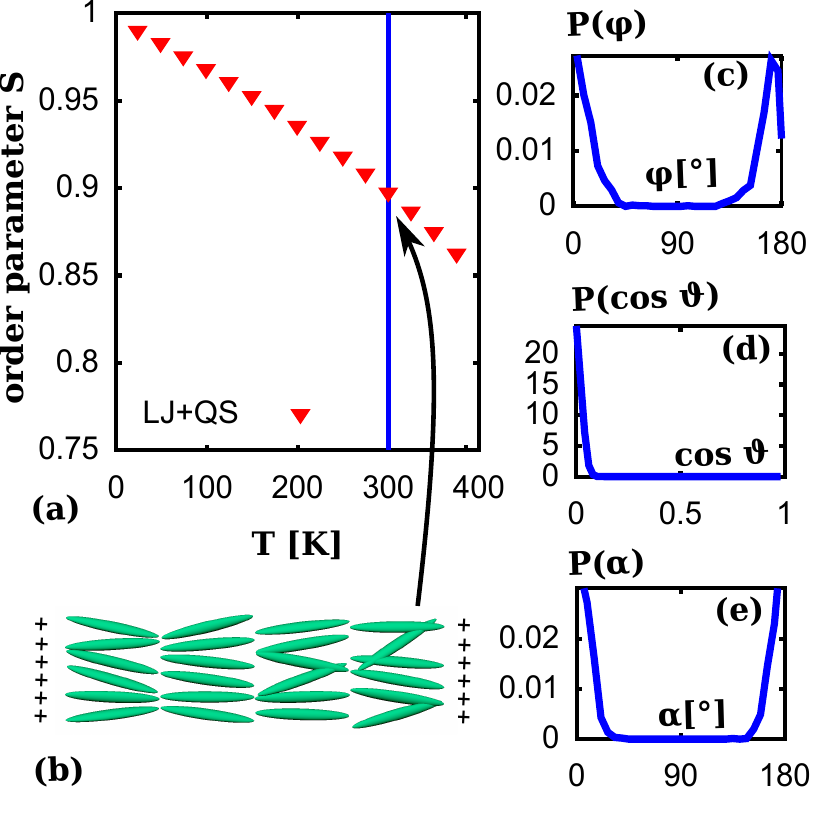}
\caption{(Color online) Ordering of a system of non-interacting molecules subject to the full substrate potential [see Eq.~(\ref{eq:mol_subs})]. 
Part~(a) depicts the nematic order parameter $S$ as function of temperature. Part~(b) contains a snapshot of a subset of the molecules at 300K, represented through the vertical line in part (a), 
while parts (c)-(e) contain corresponding angular distributions.}
\label{fig:12_temperature_subs}
\end{figure}

We now discuss the impact of the electrostatic field generated by a ZnO(10-10) surface. 
Based on our considerations in section~\ref{Sec:2B3_electr_field}, where we constructed the corresponding molecule-substrate potential [see Eq.~(\ref{eq:mol_subs})], 
we expect an individual molecule 
to lie flat on the substrate and to align with the line charges, if the temperature is sufficiently low. To determine the degree of substrate-induced ordering at the temperature studied here
($T=300K$), we consider in Fig.~\ref{fig:12_temperature_subs}(a) the order parameter $S$ of a system of {\it non-interacting} molecules as function of $T$.
It is seen that the substrate alone induces a very large degree
of single-particle ordering ($S\approx 0.9$) even at room temperature. Upon cooling the system, $S$ further increases as one would expect. 
We recall in this context that, within our model, the ordering is mediated through the interaction of the substrate field
with the fictitious quadrupole moment ${\bf q}_i'$, which lies parallel to the molecular $y'$ axes. Thus, for perfect substrate-induced order, 
all ${\bf q}_i'$ lie parallel to the $y$-axis of the coordinate system. 
Figures \ref{fig:12_temperature_subs}(b)-(e) show a snapshot of the (decoupled) many-particle system and corresponding 
angular distribution functions, respectively, at $T=300K$. 

\begin{center}
\begin{figure*}
\begin{minipage}{17cm}
    \includegraphics[bb=0 0 500 250]{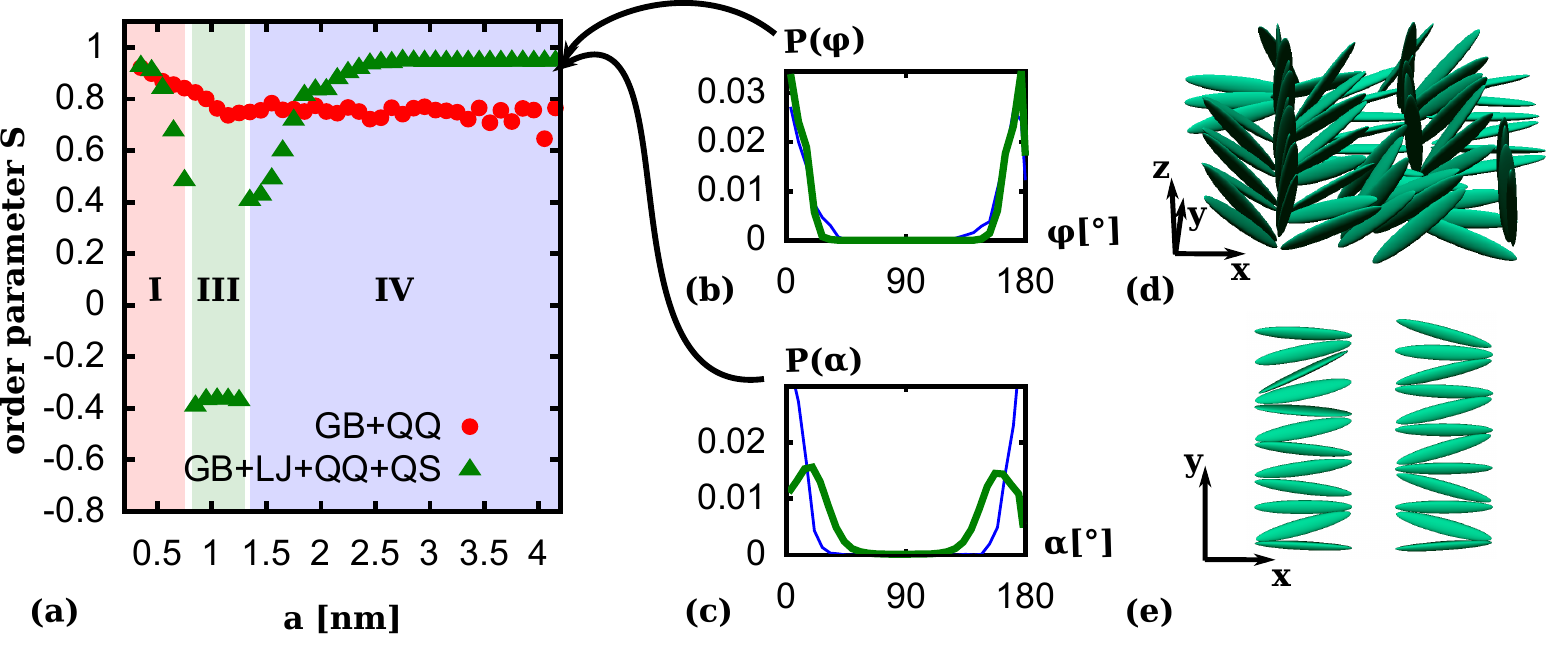}
\caption{(Color online) (a) Nematic order parameter of the fully interacting system including all types of molecule-molecule and molecule-substrate interactions. The different ordering states are denoted through roman numerals: I upright uniaxial order, 
III disorder within the $x$-$z$-plane, and IV planar uniaxial order in the $x$-$y$-plane.
As a reference, we have included data for the system without molecule-substrate interaction (see Fig.~\ref{fig:10_tetra_GB}).
Parts (b) and (c) present the distributions for the angles $\varphi$ and $\alpha$, respectively, at 
$a=4.0$~nm. The thick green lines in parts (b)-(c) represent the 
angular distributions for the system including all types of interactions, while the thin blue lines represent a system with only 
molecule-substrate interaction [see Figs. \ref{fig:12_temperature_subs}(c) and (e)]. Parts (d) and (e) show snapshots at $a=1.0$~nm (phase III)
and $a=4.0$~nm (phase IV).}
\label{fig:13_tetra_all}
\end{minipage}
\end{figure*}
\end{center}

In Fig.~\ref{fig:13_tetra_all}(a) we present the order parameter $S$ of a system with the same strength of molecule-substrate interaction, 
but full (GB plus quadrupolar) molecule-molecule interactions, as function 
of $a$ (tetragonal lattice). 
We find that the substrate-induced ordering ``survives" only at very large $a$, that is, at very low densities, whereas the high-density behavior
is dominated by molecule-molecule interactions.
Specifically, for $a\lesssim 0.6$ nm we recover the upright-uniaxial phase (I) with herringbone-like order of the molecular quadrupoles ${\bf q}_i$. 
In the range
$0.7$ nm $\lesssim a \lesssim $ $1.3$~nm, 
the system displays a state resembling state III discussed before [see Fig.~\ref{fig:11_tetra_GB_part2}(b)], that is, 
the majority of molecules lies within the $x$-$z$ plane [see Fig.~\ref{fig:13_tetra_all}(d)]. 
However, the range of corresponding lattice constants is smaller here. 
Rather we observe $S$ to {\it rise} again already at $a\approx 1.4$~nm.
This suggests that the molecule-substrate potential, which favors planar uniaxial order, now starts to ``outwin" the 
impact of the intermolecular interactions. Finally, the system enters a state of nearly perfect planar uniaxial order, as illustrated by the angular probability distributions
plotted in Figs.~\ref{fig:13_tetra_all}(b) and (c). The only difference to the non-interacting system studied before appears when we consider the function $P(\alpha)$: In the system with electrostatic substrate field,
the peaks occur at $\alpha \approx 24^{\circ}$ and $\alpha \approx 156^{\circ}=180^{\circ}-24^{\circ}$, thus they are somewhat shifted relative to the corresponding peaks in 
the non-interacting case, see Fig.~\ref{fig:12_temperature_subs}(e). We interpret this shift as a competition between the quadrupolar
molecule-molecule interactions, which favor
in-plane, T-like configuration of the moments ${\bf q}_i$, and the electrostatic molecule-substrate interaction. The latter favors the fictitious 
quadrupoles ${\bf q}_i'$ to lie parallel to $y$-axis, and thus counteracts the molecular quadrupole interactions.

\section{Conclusions}\label{Sec:4Conclusions}
In this paper we have proposed a classical, coarse-grained model for systems of 6P molecules at electrostatically patterned
(ZnO 10-10) surfaces. Within our model, the molecules are represented by rigid ellipsoids with
point multipole moments motivated by the charge distribution of true 6P molecules. Likewise, the 
surface is considered as an external perturbation giving rise to a static electric field. 

Thus, our model lacks of atomistic degrees of freedom and, thus, the ability to respond to the adsorption process by conformational changes. However, it does take into 
account electrostatic contributions to both, molecule-molecule and molecule-substrate interactions. 

Based on our coarse-grained Hamiltonian, we have performed MC simulations of a many-particle system focusing on the orientational ordering as function of density and lattice type.
In the absence of the surface pattern the model predicts, in addition to (3D) disordered states, standing and lying uniaxial configurations, 
which have also been observed in experiments \cite{Blumstengel2010,Resel2008}. Within these ordered phases, the quadrupole-quadrupole interactions play a decisive role for the nearest-neighbor 
configurations; in particular, we observe structures resembling that in a herringbone
lattice characteristic of bulk 6P systems \cite{Socci1993,Resel2008}. The electrostatic substrate field then causes pronounced alignment of the molecules along the charge lines of the 
Zno(10-10) surface for a wide range of densities. Only at the hightest density considered, 
the system with electrostatic substrate field still displays an upright uniaxial phase.

Clearly, our model neglects various, potentially important features, in particular the molecular flexibility. The latter implies that the molecule's charge distribution 
(and that of the surface!) can change
during the adsorption process, yielding a change of strength (and even nature) of the resulting multipole moments. For example, it has been suggested 
\cite{Sala2011}
that the interaction of the ZnO substrate with a 6P molecule yields an induced molecular dipole moment, which is absent when considering the same molecule in vacuum.
Nevertheless, the fact that in our rigid model we observe similar phases as in experiments suggests that we have captured key ingredients of this complex HIOS system.

Moreover, one main advantage of our model is that we can easily include additional features such as an additional (permanent) dipole, which would arise when
one considered 6P derivatives \cite{Garmshausen2014, Salzmann2008}. Even with such additions, the model is still sufficiently simple to allow for large-scale simulations inaccessible in full 
all-atom MD
simulations. We thus consider our model as a useful starting point for more elaborate coarse-grained simulations. 
An immediate and important extension of the present study would be the inclusion of translational degrees of freedom in the MC simulations. This would allow, in particular, 
to explore the stability of the various high-density phases observed in our study.

\section{Acknowledgements}\label{Sec:5Acknows}
We gratefully acknowledge stimulating discussions
with F.~Henneberger, S.~Blumstengel, J.~Dzubiella, K.~Palczynski, A.~Zykov, M.~Sparenberg, and S.~Kowarik. 
Moreover, we thank the anonymous referees of this paper for their comments, from which we have greatly benefited.
This work was supported by the Deutsche Forschungsgemeinschaft within the framework
of the Collaborative Research Center CRC 951 (project A7).

\appendix
\section{Multipole moments}\label{App:multipole_table}

In this appendix we provide numerical values for the multipole moments defined in Eqs.~(\ref{Eq:Monopole})-(\ref{Eq:Hexadec}), see Table~\ref{tab:2Multipoles}.
\begin{center}
\begin{table*}
\begin {tabular}{c|c c c}
 \hline \hline
&\qquad planar \qquad&\qquad twisted\qquad&\qquad Della Sala \textit{et al.} \qquad\\
\hline
 $q$ [e]\qquad&\quad 0.004 \qquad&\qquad 0 \qquad&\qquad - \qquad\\
\hline
 $p_{x}, p_{y}, p_{z}$ [e\,nm]\qquad&\quad 0.003, 0, 0 \quad&\quad -0.0002, 0, 0\qquad&\qquad - \qquad\\
\hline
 $Q_{xx}, Q_{yy}, Q_{zz}$ [e\,nm$^2$]\qquad &\quad  0.015, 0.011, -0.026 \qquad &\quad  0.032, 0.039, -0.071\quad & \quad 0.015, 0.013, -0.028 \,\\
\hline
 $O_{xxx}, O_{xxy}, O_{xxz}$ [e\,nm$^3$]&\quad 0.028, 0, 0 \quad&\quad -0.0005, 0.0007, 0\qquad& - \\
 $O_{yyx}, O_{yyy}, O_{yyz}$ [e\,nm$^3$]&\quad 0.007, 0, 0 \quad&\quad 0.0003, -0.0006, 0\qquad& - \\
 $O_{zzx}, O_{zzy}, O_{zzz}$ [e\,nm$^3$]&\quad -0.035, 0, 0 \quad&\quad 0.0002, -0.0001, 0\qquad& - \\
\hline
 $H_{xxxx}, H_{yyyy}, H_{zzzz}$ [e\,nm$^4$]&\quad0.432, -0.087, 0.024 \quad&\quad1.119, -0.242, 0\qquad& - \\
 $H_{xxyy}, H_{xxzz}, H_{yyzz}$ [e\,nm$^4$]&\quad0.037, -0.060, -0.053 \quad&\quad0.098, -0.171, -0.182\qquad& - \\
 $H_{yyxx}, H_{zzxx}, H_{zzyy}$ [e\,nm$^4$]&\quad0.069, 0, -0.015 \quad&\quad 0.186, 0, -0.045\qquad& - \\
 $H_{xxxy}, H_{xxxz}, H_{yyyx}$ [e\,nm$^4$]&\quad0, 0, 0\quad&\quad -0.061, 0, -0.053\qquad& - \\
 $H_{yyyz}, H_{zzzx}, H_{zzzy}$ [e\,nm$^4$]&\quad0, 0, 0\quad&\quad 0, 0, 0\qquad& - \\
 \hline \hline
\end {tabular}
\caption {Multipole moments related to different atomistic configurations of a 6P molecule. 
Specifically, the table includes values for the monopole $q$, dipole $p_{\alpha}$, quadrupole $Q_{\alpha \beta}$, 
octupole $O_{\alpha \beta \gamma}$
and hexadecapole moment $H_{\alpha \beta \gamma \delta}$ for a planar and a twisted molecular configuration. Included are results
from the DFT study of Della Sala \textit{et al.} \cite{Sala2011}. The corresponding values $Q_{\alpha\beta}$ follow from the 
values $M_{\alpha\beta}$ given in \cite{Sala2011} by using the relation \cite{priv_Blumstengel} $Q_{\alpha \beta}=(M_{\alpha \beta}-1/3 \sum_{\alpha} M_{\alpha\alpha})/2$. All multipole moments are listed in their eigensystem (which eliminates many elements). Also note 
the following symmetries: $O_{\alpha \alpha \beta}=O_{\alpha \beta \alpha}=O_{ \beta\alpha \alpha}$, $H_{\alpha\alpha\beta\beta}=H_{\alpha\beta\alpha\beta}=H_{\beta\alpha\alpha\beta}$ and 
$H_{\alpha\alpha\alpha\beta}=H_{\alpha\alpha\beta\alpha}=H_{\alpha\beta\alpha\alpha}=H_{\beta \alpha \alpha\alpha}$.}
\label{tab:2Multipoles}
\end{table*}
\end{center}

\section{Energy considerations for two linear quadrupoles}\label{App:QQ_configs}
\begin{figure}
    \includegraphics[bb=0 0 300 100]{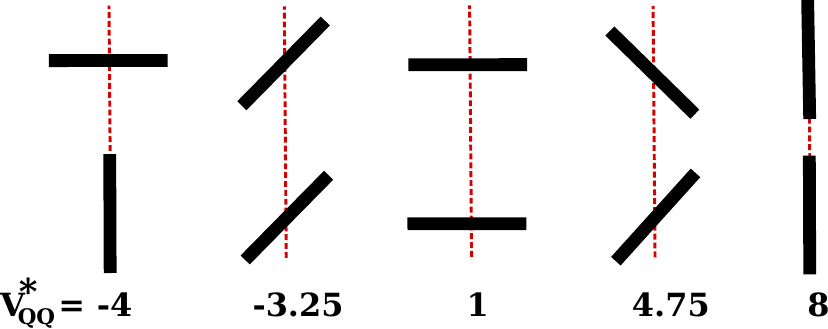}
\caption{(Color online) Interaction energy of two linear point quadrupoles for specific configurations.}
\label{fig:14_app_QQ_configs}
\end{figure}
In this appendix, we consider the most relevant configurations of two linear quadrupoles by studying their 
pair energy. Specifically, we consider the dimensionless expression [compare with Eq.~(\ref{eq:V_QQ})] 
\begin{align}
 \notag V_{QQ}^{\ast}(\textbf{q}_i, \textbf{q}_j, \textbf{r}_{ij})&=[1-5\cos\beta_i^2\\
\notag &-5\cos\beta_j^2-15 \cos\beta_i^2\cos\beta_j^2\\
 &+2(\cos\gamma_{ij}-5\cos\beta_i\cos\beta_j)^2]\textrm{,}\label{eq:V_QQ_2}
\end{align}
where $\beta_i$, $\beta_j$ and $\gamma_{ij}$ are defined below Eq.~(\ref{eq:V_QQ}). In Fig.~\ref{fig:14_app_QQ_configs} we show five relevant configurations together with the corresponding
values of $V_{QQ}^{\ast}$. It is seen that the T-configuration has the smallest interaction energy, closely followed by a configuration
in which both quadrupoles are parallel and form an angle of $45^{\circ}$ with the connection vector.  The interaction energy is maximal for parallel quadrupoles
that are aligned with respect to their connection vector.

\end{document}